\documentclass{article}
\usepackage{graphicx,amsmath,amssymb}
\usepackage{bm}
\usepackage{booktabs}
\usepackage{mathrsfs}
\usepackage[super,square,comma,compress]{natbib}
\usepackage{xcolor}

\begin{document}
     DOI: 10.1002/ \\
\textbf{Article type}: Review \\
\title{Thermal engineering in low-dimensional quantum devices: a tutorial review of nonequilibrium Green's function methods}
\author{Xiaobin Chen$^*$, Yizhou Liu$^{*}$, Wenhui Duan$^{*}$}
\date{}
{\let\newpage\relax\maketitle}
Dr. Xiaobin Chen\\
School of Science, Harbin Institute of Technology, Shenzhen 518055, China\\

Yizhou Liu, Prof. Wenhui Duan\\
State Key Laboratory of Low-Dimensional Quantum Physics, Department of Physics, Tsinghua University, Beijing 100084, China\\
Collaborative Innovation Center of Quantum Matter, Beijing 100084, China\\
Institute for Advanced Study, Tsinghua University, Beijing 100084, China\\

Email:chenxiaobin@hit.edu.cn, liu-yz12@mails.tsinghua.edu.cn, duanw@tsinghua.edu.cn\\

\textbf{Keywords:}
Phonon NEGF, thermal engineering, thermoelectricity, spin caloritronics, valley caloritronics

\begin{abstract}
Thermal engineering of quantum devices has attracted much attention since the discovery of quantized thermal conductance of phonons. Although easily submerged in numerous excitations in macro-systems, quantum behaviors of phonons manifest in nanoscale low-dimensional systems even at room temperature. Especially in nano transport devices, phonons move quasi-ballistically when the transport length is smaller than their bulk mean free paths. It has been shown that phonon nonequilibrium Green's function method (NEGF) is effective for the investigation of nanoscale quantum transport of phonons. In this tutorial review two aspects of thermal engineering of quantum devices are discussed using NEGF methods. One covers transport properties of pure phonons; the other concerns the caloritronic effects, which manipulate other degrees of freedom, such as charge, spin, and valley, via the temperature gradient. For each part, we outline basic theoretical formalisms first, then provide a survey on related investigations on models or realistic materials. Particular attention is given to phonon topologies and a generalized phonon NEGF method. Finally, we conclude our review and summarize with an outlook.
\end{abstract}

\setcounter{tocdepth}{2}
\tableofcontents

\section{Introduction}
    Rapid developments of nano devices and experimental techniques have sparked both research interests and urgent needs in thermal engineering of low-dimensional devices.\cite{LiBaowen_RMP_2012,ZhangGang_book,Cahill_JAP_2003} As the smallest lateral feature sizes of devices approach 10 nm, the chip-level power density reaches 10$^2$ W/cm$^2$, which is comparable to that of a nuclear reactor.\cite{Pop_IEEE_2006} Without efficient heat dissipation, such high power density is detrimental to the stability, reliability, and performance of nano devices. Hence, it is necessary to investigate the thermal transport properties of nano structures based on silicon and other materials which are promising for replacing silicon in the next-generation of electronics.\cite{Graham_small2005} This demand has created a significant amount of theoretical and experimental research on low-dimensional systems, which not only show great potential for realistic applications, but also provide a platform for investigating fundamental physics.

    From the aspect of realistic applications, materials with high thermal conductance/conductivity are needed for heat dissipation in highly integrated electronic circuits, but those with low thermal conductance/conductivity are favorable for thermoelectric applications. In fact, thermal conductivity of low-dimensional materials is much more wide-ranged than those of three-dimensional (3D) bulk materials.\cite{LiBaowen_AdvMat2013} For example, graphene has exceedingly high thermal conductivity of up to about $\sim 5 \times 10^3$ W/m$\cdot$K, which is more than two times larger than that of natural diamond.\cite{Balandin_NanoLett2008,Xuyong_small_2014} By constrast,  thermal conductivities of silicon nanowires with rough surfaces can be reduced by 100 times compared to that of bulk silicon.\cite{YaPeidong_2008,Majumda}

    From the aspect of fundamental physics, low-dimensional systems display various quantum behaviors of phonon transport which are drastically distinct from the classical ones. The first is the breakdown of the classical Fourier's law of heat conduction, which indicates that the heat flux density driven by a temperature gradient is determined by the geometry-independent thermal conductivity. In 2008, Chang \emph{et~al.} presented evidence for the breakdown of the classical Fourier's law in quasi-1D materials, such as multiwalled carbon and boron-nitride nanotubes, where thermal conductivities depend on the length of nanotubes.\cite{Zettl_PRL_2008} Size dependence of thermal conductivity was also shown in 2D materials afterwards.\cite{LiBaowen_NatComm_2014} The second is the quantum ballistic transport in low-dimensional systems, which is different from the classical diffusive transport in 3D bulk materials.
    Thermal conductance was predicted to be quantized at low temperatures in the ballistic phonon regime.\cite{Pendry_1983,Maynard_PRB_1985,Angelescu_1998,Rego_PRL_1998,Blencowe_PRB_1999}
    This prediction was convinced in 2000, when Schwab \emph{et~al.} experimentally observed the quantum of thermal conductance in suspended silicon nitride membrane.\cite{Schwab_nat_2000} The last, but not the least, is about quantum confinement of phonons. In 2000, Hone \emph{et~al.} showed indications of a quantized 1D phonon spectrum in single-walled carbon nanotubes (CNTs).\cite{HoneJ_Sci_2000} Also in 2005, significant phonon confinement effects in silicon nanowires down to 4 nm in diameter were reported using Raman microscopy studies.\cite{Adu_NanoLett2005}

     Among this research, of particular interest is thermal engineering in low-dimensional quantum devices, which can be classified into two classes: (i) manipulating phonons to engineer phonon conduction or design phonon circuits;\cite{LiBaowen_RMP_2012} (ii) manipulating other degrees of freedom, such as charge, spin,\cite{Breton_Nat_2011,cxb_PRB_2014,XiaoJiang_PRB2010,Cheng_PRL2016,TangGaomin_arxiv2017} and valley,\cite{cxb_PRB2015valley} with the aid of temperature gradient. Nonequilibrium Green's function method is an effective tool for studying the quantum thermal engineering of nanoscale systems. In this review, we aim to give a detailed summary of the phonon nonequilibrium Green's function (NEGF) method and the applications of the NEGF method for electrons on thermal engineering of quantum devices. First, we briefly introduce the phonon NEGF method and its applications, then present a short discussion on generalizing the phonon NEGF method for topological phonon devices. Second, we illustrate how basic concepts of thermoelectricity generally apply to valley and spin caloritronics.
     Finally, we conclude with an outlook. To further stimulate the research interest in this field and make this review appeal to a broader audience, we describe the formalism as detailedly as possible. People who are already familiar with the phonon NEGF method can directly skip Section 2.1 to other parts.

\section{Phononic devices}

    \subsection{Phonon NEGF Method}
     Boltzmann transport equation method (BTE)\cite{YangRonggui_PRB2004,LiWu_PRB2015,LiWu_CPC2016} and molecular dynamics (MD) method\cite{LiBaowen_RMP_2012, ZhangGang_JCP_2005,chenYP_NanoLett_2009,GongXG_APL2009, Keblinski_APL2010, Pop_PRB2011, ZhangGang_Nanoscale2011, Cagin_ACSNano2011,  XuBaoxing_Carbon2016, ChenChaoguang_Carbon2017} are two effective methods for studying thermal transport. The BTE method deals with diffusive transport and provides accurate thermal conductivity for bulk systems. The classical MD method naturally incorporates nonlinearity and works at high temperatures where classical behaviors dominate.
     For low-dimensional systems where quantum effects become important,
     the phonon NEGF method\cite{WangJS_EPJ_2008, Xuyong_APL2009, Xuyong_prb_2010, ZhuHongqin_NJP2012, cxb_PUCNT, Huaqing_PRB_2013, cxb_PRB_2014} is  effective in a whole diffusive-to-ballistic regime.
     There are also other methods, such as continuum models for elastic waves,\cite{WangJS_EPJ_2008,ChenKeqiu_SciRep2015} and hydrodynamic models based on thermomass theory.\cite{GuoZengyuan2015}
     In the following, we give a basic formalism of the phonon NEGF method and introduce its application in the study of low-dimensional systems.

        \subsubsection{Hamiltonian of a harmonic-vibrating system}
        The Hamiltonian of a vibrating system can generally be written into two parts\cite{Ziman_book}
        \begin{align}
        H = T + V,
        \end{align}
        where $T$ is kinetic energy, and $V$ is the potential energy.
        Expanding the potential in powers of displacements of atoms around the equilibrium positions, we have
        \begin{align}\label{eq:H}
        & H\left( {\left\{ {{{\bf{R}}_I}} \right\}_{I = 1}^N} \right) \cr
        & \approx \sum\limits_{I = 1}^N {\sum\limits_{\alpha  = xyz}^{} {\frac{{p_{I\alpha }^2}}{{2{M_I}}}} }  + \frac{1}{2}\sum\limits_{I\alpha ,J\beta }^{} {{{\left. {\frac{{{\partial ^2}V}}{{\partial {R_{I\alpha }}\partial {R_{J\beta }}}}} \right|}_0}{\eta _{I\alpha }}{\eta _{J\beta }}} ,\quad
        \end{align}
        where $\left\{ {{{\bf{R}}_I}} \right\}_{I = 1}^N$ is an assembly of atomic positions, ${{\bf{R}}_I}$ is the position vector of the $I^{\textrm{th}}$ atom, $p_{I,\alpha }^{}$ is the momentum of the $I^{\textrm{th}}$ atom along the $\alpha$ axis ($\alpha=x,y,z$), and ${\eta _{I\alpha }} = {R_{I\alpha }} - R_{I\alpha }^0$ indicates the displacement with respect to the equilibrium position of the $I^{\textrm{th}}$ atom. The constant energy term is omitted, which does not influence the dynamics of the system. The first-order term of $\eta_{I\alpha}$ is zero at the equilibrium position. In order to reduce the information about atomic masses, one may introduce mass-weighted displacements, momenta, and the dynamical matrix as
        \begin{align}
        {u_{I\alpha }} &\equiv \sqrt {{M_I}} {\eta _{I\alpha }},\quad {v_{I\alpha }} \equiv {\dot u_{I\alpha }} = {p_{I\alpha }}/\sqrt {{M_I}}, \\
        {D_{I\alpha ,J\beta }} &\equiv \frac{1}{{\sqrt {{M_I}{M_J}} }}{\left. {\frac{{{\partial ^2}V}}{{\partial {R_{I\alpha }}\partial {R_{J\beta }}}}} \right|_0},
        \end{align}
        and the original Hamiltonian in Equation~(\ref{eq:H}) can be written in a compact form as
         \begin{align}
         H\left( {\left\{ {{{\bf{R}}_I}} \right\}_{I = 1}^N} \right) = \sum\limits_{I = 1}^N {\sum\limits_{\alpha  = xyz}^{} {\frac{1}{2}} } v_{I\alpha }^2 + \frac{1}{2}\sum\limits_{I\alpha ,J\beta }^{} {{D_{I\alpha ,J\beta }}{u_{I\alpha }}{u_{J\beta }}} .
         \end{align}
         The displacements and momenta operators satisfy
        \begin{align}
         [{\hat\eta _{I\alpha }},{\hat p _{J\beta }}] &= i\hbar \delta_{IJ} \delta_{\alpha\beta},\\
        [{\hat \eta _{I\alpha }},{\hat\eta _{J\beta }}] &= 0,\;\quad [\hat p_{I\alpha }^{},\hat p_{J\beta }^{}] = 0,
        \end{align}
        then the mass-weighted ones have a similar relation as
        \begin{align}
        [\hat u_{J\beta }^{}, \hat v_{I\alpha }^{}] &= [\sqrt {{M_J}}\hat \eta _{J\beta }^{},\frac{{{\hat p}_{I\alpha }^{}}}{{\sqrt {{M_I}} }}] = {\textrm i} \hbar {\delta _{IJ}}{\delta _{\alpha \beta }}, \\
        [{\hat u_{I\alpha }},{\hat u_{J\beta }}] &= 0,\;\quad [\hat v_{I\alpha }^{},\hat v_{J\beta }^{}] = 0.
        \end{align}

        \subsubsection{Thermal current in terms of Green's function}
\begin{figure}
         \centering
      \includegraphics[width=0.4\textwidth]{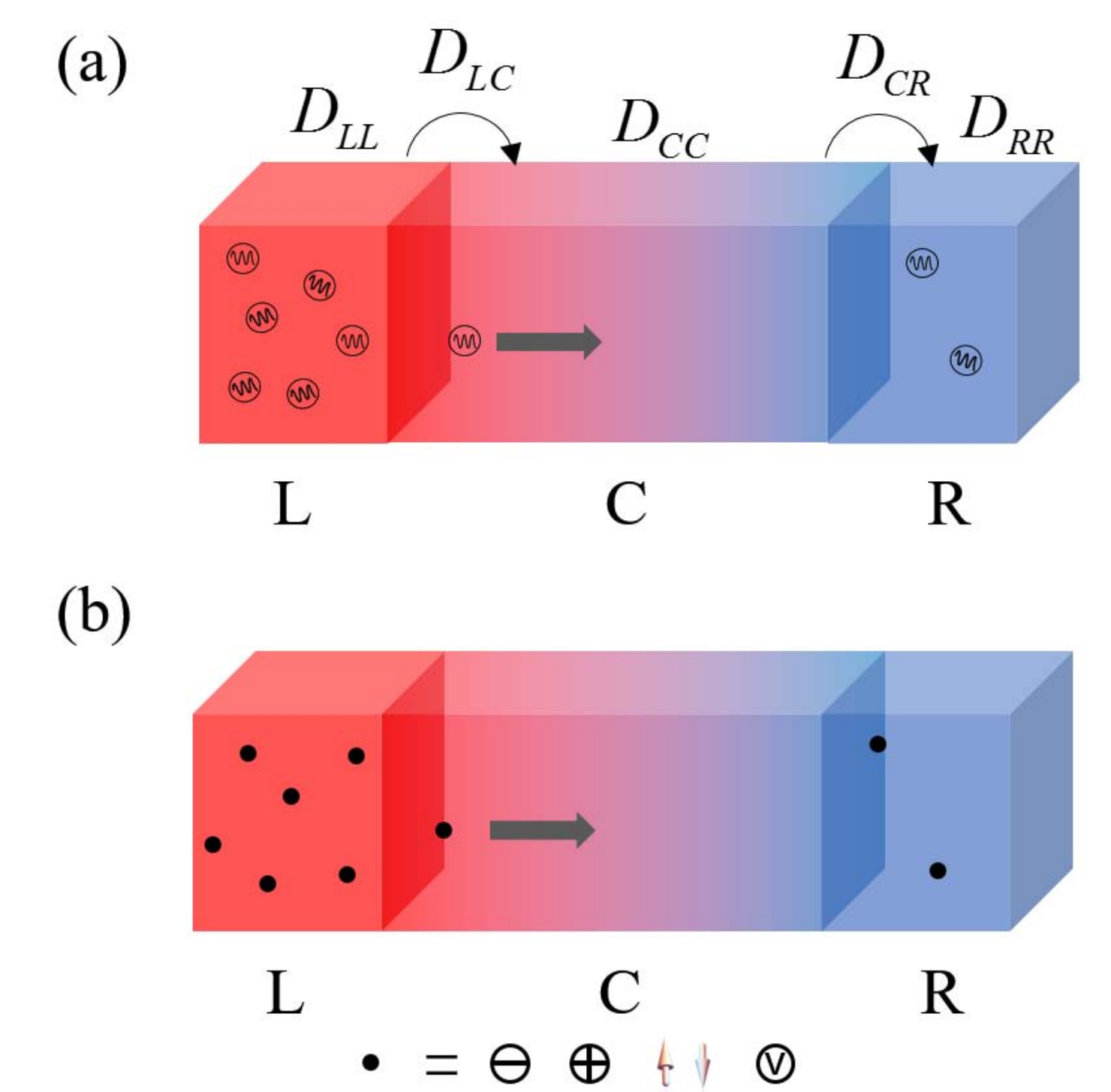}\\
      \caption{Thermal engineering in quantum devices. Temperature gradient drives the transport of (a) phonons in a phononic device, and (b) other degrees of freedom, such as charge, spin, and valley, in a caloritronic device. Particles or quasi-particles that comes from the source ($L$) transport quasi-ballistically to the drain ($R$) when the size of the device is smaller than the bulk mean free path. }\label{fig:LCR}
    \end{figure}

        {\bf{Thermal current.}} For a two-probe system consisting of the left ($L$) and right ($R$) thermal leads, connected by a long central region (Figure 1) so that the interaction between two leads become negligible, the Hamiltonian can be expressed as\cite{WangJS_EPJ_2008}
        \begin{align}
        \hat H = {\hat H_L} + {\hat  H_R} + {\hat H_C} + {\hat V_{LC}} + {\hat V_{CR}}
        \end{align}
        with
        \begin{align}
        {\hat H_{L/R/C}}& = \sum\limits_{I \in {N_{L/R/C}}}^{} {\sum\limits_{\alpha  = xyz}^{} {\frac{1}{2}} } \hat v_{I\alpha }^2 \cr
        &+ \frac{1}{2}\sum\limits_{I,J \in {N_{L/R/C}}  } {\sum\limits_{\alpha ,\beta  = xyz}^{} {{\hat u_{I\alpha }}{D_{I\alpha ,J\beta }}{\hat u_{J\beta }}} } ,\\
        {\hat V_{LC}} &= \frac{1}{2}\left( {\sum\limits_{I \in L \atop J \in C}^{}  +  \sum\limits_{I \in C \atop J \in L}^{} {} } \right)\sum\limits_{\alpha ,\beta }^{} {{\hat u_{I\alpha }}{D_{I\alpha ,J\beta }}{\hat u_{J\beta }}} ,\\
        {\hat V_{CR}} &= \frac{1}{2}\left( {\sum\limits_{I \in R \atop J \in C}^{}  +  \sum\limits_{I \in C \atop J \in R}^{} {} } \right)\sum\limits_{\alpha ,\beta }^{} {{\hat u_{I\alpha }}{D_{I\alpha ,J\beta }}{\hat u_{J\beta }}} .
        \end{align}
        When written in the matrix form, the dynamical matrix of the system is [see Figure~\ref{fig:LCR}(a)]
        \begin{align}
        {\bf D} = \left( {\begin{array}{*{20}{c}}
        {{{\bf D}_{LL}}}&{{{\bf D}_{LC}}}&{\bf{0}}\\
        {{{\bf D}_{CL}}}&{{{\bf D}_{CC}}}&{{{\bf D}_{CR}}}\\
        {\bf{0}}&{{{\bf D}_{RC}}}&{{{\bf D}_{RR}}}
        \end{array}} \right)
        \end{align}
        with dimensions of each block indicated by
        \begin{align}
        \left( {\begin{array}{*{20}{c}}
        {s{N_L} \times s{N_L}}&{s{N_L} \times s{N_C}}&{s{N_L} \times s{N_R}}\\
        {s{N_C} \times s{N_L}}&{s{N_C} \times s{N_C}}&{s{N_C} \times s{N_R}}\\
        {s{N_R} \times s{N_L}}&{s{N_R} \times s{N_C}}&{s{N_R} \times s{N_R}}
        \end{array}} \right).
        \end{align}
        Here, $s$ stands for the number of oscillating degrees of freedom for each atom, and $N_{L/C/R}$ is the number of atoms in the $L/C/R$ region. Normally, $L$ and $R$ are supposed to be phonon reservoirs, having infinite degrees of freedom, \textit{i.e.}, $N_{L/R}=+\infty$.

        The phonon current flowing out of a thermal lead, such as lead $L$, can be measured by the variation of energy in the thermal lead:\cite{wang_prb_2006}
        \begin{align}\label{eq:JL=-dHL}
        {J_L}\left( t \right) = \sum\limits_{I \in L}^{} {{J_I}\left( t \right)}  =  - \left\langle {{{\frac {{\textrm d}\hat  H_L}{{\textrm d} t}}}} \right\rangle .
        \end{align}
        The change rate of $\hat H_L$ can be evaluated by utilizing the Heisenberg equation:
        \begin{align}\label{JL0}
        {J_L}\left( t \right) &=  - \frac{1}{{{\textrm i} \hbar }}\left\langle {[{\hat  H_L},{\hat  V_{LC}}]} \right\rangle \cr
           &= \frac{1}{2}\mathop {\lim }\limits_{t' \to t} \frac{\partial }{{\partial t'}}\sum\limits_{\scriptstyle I \in L,J \in C \atop \scriptstyle \alpha ,\beta }^{} {\left[ {{D_{I\alpha ,J\beta }}\left\langle {{\hat u_{I\alpha }}\left( {t'} \right){\hat  u_{J\beta }}\left( t \right)} \right\rangle }\right. } \cr
           &{\left. {+ {D_{J\beta ,I\alpha }}\left\langle {{\hat u_{J\beta }}\left( t \right){\hat u_{I\alpha }}\left( {t'} \right)} \right\rangle } \right]}.
        \end{align}
        This formula can be rewritten in the form of Green's functions as
        \begin{align}\label{eq:JLt}
        {J_L}\left( t \right) &= \frac{1}{2}\mathop {\lim }\limits_{t' \to t} \frac{\partial }{{\partial t'}}\left[ {\sum\limits_{I \in L,J \in C,\alpha ,\beta }^{} {{\textrm i} \hbar {D_{I\alpha ,J\beta }}G_{J\beta ,I\alpha }^ < \left( {t,t'} \right) }} \right. \cr
        &\left.  + {\textrm i} \hbar {D_{J\beta ,I\alpha }}G_{I\alpha ,J\beta }^ < \left( {t',t} \right) \right]\cr
        & = \frac{1}{2}\mathop {\lim }\limits_{t' \to t} \frac{\partial }{{\partial t'}}{\rm{Tr}}\left[ {{\textrm i} \hbar {{\bf D}_{LC}}{\bf G}_{CL}^ < \left( {t,t'} \right) + h.c.} \right],
        \end{align}
        where ``$h.c.$" is short for Hermitian conjugate, and the lesser Green's function is defined as\cite{yamamoto_PRL2006}
        \begin{align}
        G_{J\beta ,I\alpha }^ < \left( {t,t'} \right) \equiv  - \frac{\textrm i} {\hbar }\left\langle {\hat u_{I\alpha }^{}\left( {t'} \right){\hat u_{J\beta }}\left( t \right)} \right\rangle .
        \end{align}
        The bracket $\langle\cdot\rangle=\textrm{Tr}[\rho\cdot]$ averages over nonequilibrium density matrix $\rho$, and the operators ${\hat u_{I\alpha /J\beta }}\left( t \right)$ are in the Heisenberg representation. For steady state, the system is time-translational invariant, which requires that
        \begin{align}G\left( {t,t'} \right) = G\left( {t - t'} \right).\end{align}
        Thus, Fourier transform can be carried out to get
        \begin{align}
        G_{J\beta ,I\alpha } \left( {t,t'} \right) &= G_{J\beta ,I\alpha } \left( {t - t'} \right) \cr
        &= \int_{ - \infty }^{ + \infty } {G_{J\beta ,I\alpha } \left( \omega  \right){e^{ - {\textrm i}\omega \left( {t - t'} \right)}}\frac{{{\textrm d}\omega }}{{2\pi }}} .
        \end{align}
        Using the Fourier transform of the lesser Green's function in Equation~(\ref{eq:JLt}), we have
        \begin{align} \label{eq:JL}
        {J_L}\left( t \right) =  - \frac{1}{2}\int_{ - \infty }^{ + \infty } {\frac{{{\textrm d}\omega }}{{2\pi }} \cdot \hbar \omega  \cdot {\rm{Tr}}\left[ {{{\bf D}_{LC}}{\bf G}_{CL}^ < \left( \omega  \right) + h.c.} \right]} .
        \end{align}

        In NEGF formalism, it is generally assumed that the transporting system is at equilibrium at the remote past ($t=-\infty$), and at some point it is applied with nonequilibrium Hamiltonian and nonequilibrium transport starts.\cite{Jauho_Book,WangJiansheng_PRE2007} For phonon transport, it is more reasonable to assume that the thermal leads and the central region are disconnected in the remote past, then connected to each other at some time $t_0$. In this way, both thermal leads serve as reservoirs with well-defined constant temperatures. For steady-state transport, one should focus on time region $t\gg t_0$, where time-translational invariance can be assumed and utilized.
        For simplicity, we do not introduce time-ordered and contour-ordered Green's functions here. However, both of them are extremely important for time-dependent phenomena where perturbation theory is needed.

        \textbf{Thermal current in terms of central-region quantities.} Note that Equation~(\ref{eq:JLt}) is applicable to both steady and transient problems, but Equation~(\ref{eq:JL}) is only valid for steady-state transport.
        One problem for both Equations (\ref{eq:JLt}) and (\ref{eq:JL}) is that usually the matrixes ${\bf D}_{LC}$ and ${\bf{G}}_{CL}^<$ have infinite dimensions, which makes realistic calculations difficult. Consequently, transforming the equation into an expression of matrixes with finite dimensions is necessary. For this purpose, we can apply the Dyson equation to use quantities of the central region instead:
        \begin{align}
        { {\bf G}_{CL}} = {{\bf G}_{CC}}{{\bf D}_{CL}}{{\bf g}_L},
        \end{align}
        where ${\bf G}_{CC}$ is the Green's function for the central region, and ${\bf g}_L$ is the Green's function of the left lead without attachment to the central region. According to the Langreth theorem,\cite{Jauho_Book} we further have
        \begin{align}
        {\bf G}_{CL}^ <  = {\bf G}_{CC}^r{{\bf D}_{CL}}{\bf g}_L^ <  + {\bf G}_{CC}^ < {{\bf D}_{CL}}{\bf g}_L^a,
        \end{align}
        where ``$r$'' and ``$a$'' stand for ``retarded'' and ``advanced'', respectively. Correspondingly, we get
        \begin{align} \label{eq:GCL}
        {\bf G}_{CL}^ < {{\bf D}_{LC}} &= {\bf G}_{CC}^r{{\bf D}_{CL}}{\bf g}_L^ < {{\bf D}_{LC}} + {\bf G}_{CC}^ < {{\bf D}_{CL}}{\bf g}_L^a{{\bf D}_{LC}}\cr
         &= {\bf G}_{CC}^r {\bf \Sigma} _L^ <  + {\bf G}_{CC}^ < {\bf \Sigma} _L^a,
        \end{align}
        with self-energies defined as ($l = L,R; {\gamma  = r,a, > , < }$)
        \begin{align} \label{eq:sigmal}
        {\bf \Sigma} _l^\gamma  \equiv {{\bf D}_{Cl}}{\bf g}_l^\gamma {{\bf D}_{lC}}.
        \end{align}
        Note that in calculations of realistic materials, ${\bf g}_l^\gamma$ can be effectively replaced by the surface Green's function of lead $l$ and ${{\bf D}_{Cl}}$ be replaced by the interaction between the central region and the surface layer of lead $l$.
        Substituting Equation~(\ref{eq:GCL}) into Equation~(\ref{eq:JL}), we have\cite{WangJiansheng_PRE2007}
        \begin{align}
        {J_L}\left( t \right) &=  - \int_{ 0 }^{ + \infty } \frac{{{\textrm d}\omega }}{{2\pi }} \cdot \hbar \omega  \cdot {\rm{Tr}}
        \left[ {\bf G}^r\left( \omega  \right){\bf \Sigma} _L^ < \left( \omega  \right) \right. \cr
         &+ \left. {\bf G}^ < \left( \omega  \right){\bf \Sigma} _L^a\left( \omega  \right) + h.c. \right] \label{eq:JLw1}\\
         &=  - \int_0^{ + \infty } {\frac{{{\textrm d}\omega }}{{2\pi }}}  \cdot \hbar \omega  \cdot {\rm{Tr}}\left[ {\left( {{\bf G}^r - {\bf G}^a} \right){\bf \Sigma} _L^ < } \right.\cr
         & + \left. {{\bf G}^ < \left( {{\bf \Sigma} _L^a - {\bf \Sigma} _L^r} \right)} \right] . \label{eq:JLw}
        \end{align}
        Without causing ambiguity, we have omitted the subscript ``$CC$'' of ${\bf G}^\gamma$ and shall follow this convention hereinafter. Note that the integration is now restricted to be within the range $\omega\sim [0, \infty)$ due to the symmetry of the integration kernel. The thermal current is now expressed in terms of central-region quantities through the introduction of self-energies.

        By further using the following equations\cite{Jauho_Book,WangJS_FoP2014}
        \begin{align}
        {\bf G}^r-{\bf G}^a &= {\bf G}^> - {\bf G}^<, \\
        {\bf\Sigma} _\alpha ^r - {\bf\Sigma} _\alpha ^a &= {\bf\Sigma} _\alpha ^ >  - {\bf\Sigma} _\alpha ^ <,
        \end{align}
        Equation~(\ref{eq:JLw}) can be written in the form of Meir-Wingreen formula:\cite{Jauho_Book,WangJS_FoP2014}
       \begin{align}\label{JLv3_WGreen}
       {J_L}\left( t \right){\rm{ }} &= \int_{ - \infty }^{ + \infty } {\frac{{{\textrm d}\omega }}{{2\pi }}}  \cdot \hbar \omega  \cdot {\rm{Tr}}\left[ {{\bf G}^ < \left( \omega  \right){\bf\Sigma} _L^ > \left( \omega  \right)} \right.\cr
       &\left. { - {\bf G}^ > \left( \omega  \right){\bf\Sigma} _L^ < \left( \omega  \right)} \right].
       \end{align}
       Equations (\ref{eq:JLw}) and (\ref{JLv3_WGreen}) have also been used for systems beyond the harmonic approximation.\cite{WangJiansheng_PRE2007}

        \textbf{Landauer-like equation.} The retarded and advanced Green's functions of a whole system in the frequency domain for a steady-state transport are obtained as
        \begin{align}\label{eq:Gra}
        {{\bf G}^{r,a}_{\textrm{whole}}}\left( \omega  \right) = {\left[ {{{\left( {\omega  \pm {\textrm i}0^+ } \right)}^2}{\bf I} - {\bf D}_{\textrm{whole}}} \right]^{ - 1}},
        \end{align}
        where ${\bf I}$ represents an identity matrix and $0^+$ is an infinitesimal positive number.
        The Green's function of the central region is the projection of ${{\bf G}^{r,a}_{\textrm{whole}}}$ at the central region. It can be effectively obtained by taking into account the self-energies from left and right leads:\cite{Mingo_PRB_2003GF,wang_prb_2006}
        \begin{align}\label{eq:Gcc}
        {\bf G}^{r,a} = {\left[ {{{\left( {\omega  \pm {\textrm i}0^+ } \right)}^2}{\bf I} - {\bf D}_{CC}^{} - {{\bf \Sigma} ^{r,a}}} \right]^{ - 1}},
        \end{align}
        where ${{\bf \Sigma} ^{r,a}}={\bf \Sigma}_L ^{r,a}+{\bf \Sigma}_R ^{r,a}$ is the total self-energy of thermal leads.
        The phonon bandwidth function is related to the imaginary part of self-energies,
        \begin{align} \label{eq:gammaL}
        {{\bf \Gamma} _l} = {\textrm i}\left( {{\bf \Sigma} _l^r - {\bf \Sigma} _l^a} \right).
        \end{align}
        Below we also list several other very useful equations, including the Keldysh equation,\cite{wang_prb_2006}\cite{note2}
        \begin{align}\label{eq:Gcc<}
        {\bf G}^ < \left( \omega  \right) = {\bf G}^r\left( \omega  \right){{\bf \Sigma} ^ < }\left( \omega  \right){\bf G}^a\left( \omega  \right),
        \end{align}
        and
        \begin{align}\label{eq:GGG}
        & {\bf G}^r - {\bf G}^a  = {\bf G}^r\left[ {{{\left( {{\bf G}^a} \right)}^{ - 1}} - {{\left( {{\bf G}^r} \right)}^{ - 1}}} \right]{\bf G}^a\cr
         &={\bf G}^r\left[ {{{\left( {\omega  - {\textrm i}0^+ } \right)}^2 {\bf I}} - {{\bf \Sigma} ^a} - {{\left( {\omega  + {\textrm i}0^+ } \right)}^2{\bf I}} + {{\bf \Sigma} ^r}} \right]{\bf G}^a\cr
         &= {\bf G}^r\left( {{ {\bf \Sigma} ^r} - {{\bf \Sigma} ^a}} \right){\bf G}^a \cr
         &=  - {\textrm i}{\bf G}^r{\bf \Gamma} {\bf G}^a.
        \end{align}
        In addition, the lesser self-energy for thermal lead $l$ is
        \begin{align} \label{eq:sigmaL}
        {\bf \Sigma} _l^ < \left( \omega  \right) =  - {\textrm i}{f_{BE;l}}\left( \omega  \right){{\bf \Gamma} _l}\left( \omega  \right),
        \end{align}
        where $f_{BE;l}(\omega)=1/\left[ {\exp (\hbar \omega /{k_B}T_l) - 1} \right]$ is the Bose-Einstein distribution function for Bosons in thermal lead $l$, and $k_B$ is the Boltzmann constant.
        With the help of Equations~(\ref{eq:gammaL})-(\ref{eq:sigmaL}), Equation (\ref{eq:JLw}) can be further simplified to the Landauer-like formula:\cite{Mingo_PRB_2003GF,wang_prb_2006,Meir_PRL1992}
        \begin{align}\label{eq:JLv2}
        {J_L}\left( t \right)  &= \int_0^{ + \infty } \frac{{{\textrm d}\omega }}{{2\pi }} \hbar \omega \left[ {{f_{BE;L}}\left( \omega  \right) - {f_{BE;R}}\left( \omega  \right)} \right] \cdot  \Xi_{ph}(\omega)
        \end{align}
        with phonon transmission function expressed in the form of the Caroli formula\cite{CaroliFormula_JPCS1971}
        \begin{align}\label{eq:trans}
        \Xi_{ph}(\omega)={\rm{Tr}}\left[ {{{\bf \Gamma} _L}\left( \omega  \right){\bf G}_{CC}^r\left( \omega  \right){{\bf \Gamma} _R}\left( \omega  \right){\bf G}_{CC}^a\left( \omega  \right)} \right].
        \end{align}
        This formula is only applicable to quasi-ballistic cases, where inelastic processes,
        such as electron-phonon scattering described by\cite{Jauho_PRB_1994}
        \begin{align}
        H_{C}^{\textrm{e-ph}} = {\varepsilon _0}{d^ {\dagger} }d + {d^ {\dagger} }d \sum\limits_{\bf{q}}  {M_{\bf{q}}}\left( {a_{\bf{q}}^ {\dagger}  + {a_{ - \bf{q}}}} \right)
        \end{align}
        in a single-level quantum dot with $d^ {\dagger} (d)$ creating (annihilating) an electron in the central quantum dot and $a_{\bf{q}}^ {\dagger}(a_{-{\bf{q}}})$ creating (annihilating) a phonon with wavevector $\bf{q}$($\bf{-q}$), are excluded. By using an ideal transmission function, Equation (\ref{eq:JLv2}) provides us with an upper limit for the thermal current in a given system.\cite{WangJS_EPJ_2008}

        \textbf{Local thermal current.}
        In the section above, we  have obtained the expression for total thermal current. Similarly, local thermal current can be obtained following the same procedure. Thermal current flowing out of $I^{\textrm {th}}$ atom is 
        \begin{align}
         {J_I}\left( t \right) =   - \left\langle {\frac{ {\textrm d}{\hat H}_I}{{\textrm d}t}} \right\rangle .
        \end{align}
        Using the Heisenberg equation, the steady-state thermal current can be written out as
        \begin{align}
        {J_I}\left( t \right) &=  - \frac{1}{2}\mathop {\lim }\limits_{t' \to t} \frac{\partial }{{\partial t'}}\sum\limits_{J\ne I,\alpha ,\beta }^{} {\left[ {{\textrm i} \hbar {D_{I\alpha ,J\beta }}G_{J\beta ,I\alpha }^ < \left( {t,t'} \right) + h.c.} \right]} \cr
         &=  - \sum\limits_{J\ne I \atop \alpha ,\beta }^{} {\left[ {\int_0^{ + \infty } {\frac{{{\textrm d}\omega }}{{2\pi }} \cdot \hbar \omega  \cdot {D_{I\alpha ,J\beta }}G_{J\beta ,I\alpha }^ < \left( \omega  \right)}  + h.c.} \right]} .
        \end{align}
        Here, we transfer to frequency domain to get rid of the differentiation over $t'$. Thus, the energy transferred from the $I^{\textrm {th}}$  atom along the $\alpha$ direction to the $J^{\textrm {th}}$($J\ne I$) atom on the $\beta$ direction is\cite{mingo_prb_2006,Morooka_PRB2008}
        \begin{align}
        {J_{I\alpha  \to J\beta }}\left( t \right) =  - 2\int_0^{ + \infty } {\hbar \omega {\mathop{\rm Re}\nolimits} \left[ {{D_{I\alpha ,J\beta }}G_{J\beta ,I\alpha }^ < \left( \omega  \right)} \right]\frac{{{\textrm d}\omega }}{{2\pi }}}.
        \end{align}
        The total thermal current flowing out of the $I^{\textrm {th}}$ atom to the $J^{\textrm {th}}$ atom is a summation over all directions
        \begin{align}
        {J_{I \to J}}\left( t \right) = \sum\limits_{\alpha ,\beta  = x,y,z}^{} {{J_{I\alpha  \to J\beta }}\left( t \right)},
        \end{align}
        and the total thermal flowing out of the $I^{\textrm {th}}$ atom is a summation over all atoms
        \begin{align}
        {J_{I}}\left( t \right) = \sum\limits_{J\ne I}^{} {{J_{I \to J}}}(t).
        \end{align}

        \subsubsection{Thermal conductance}
        Thermal conductance is measured by the ratio of thermal flux and temperature difference as
        \begin{align}
        K = \frac{{{J_{L \to R}}}}{{{T_L} - {T_R}}}.
        \end{align}
        Thermal conductivity ($\kappa$) is thermal conductance scaled with the cross sectional area $A$ and the transport length $L$ of a given sample:\cite{Mingo_PRB_2003GF}
        \begin{align}
        \kappa  = K \cdot \frac{L}{A}.
        \end{align}
        Thermal conductivity contributed by phonons ($\kappa_{ph}$) of low-dimensional material is highly dependent on geometry and size.\cite{LiBaowen_small2012,ZhangGang_EPJB2012} Thermal conductance ($K_{ph}$), however, remains independent of transport length in the ballistic transport regime. To see this phenomenon more clearly, we consider a system which has tiny temperature difference between the left and right leads, $\Delta T= T_L - T_R$, and expand Equation~(\ref{eq:JLv2}) to the linear order of temperature difference around the average temperature $T=(T_L+T_R)/2$ to get\cite{yamamoto_PRL2006,WangJS_EPJ_2008,Xuyong_prb_2010}
        \begin{align}
        {J_Q} &\approx \frac{1}{{2\pi }}\int_0^{ + \infty } {{{\left. {\frac{{\partial f_{BE}}}{{\partial T}}} \right|}_T} \cdot \left( {{T_L} - {T_R}} \right)}  \cdot \hbar \omega  \cdot \Xi_{ph} \left( \omega  \right) \cdot {\textrm d} \omega.
        \end{align}
        This equation leads to the expression for thermal conductance in the framework of phonon NEGF:\cite{yamamoto_PRL2006}
        \begin{align}\label{eq:thermConductance}
        K_{ph}\left( T \right)&=\frac{1}{{2\pi }}\int_0^{ + \infty } {{{\left. {\frac{{\partial f_{BE}}}{{\partial T}}} \right|}_T} }  \cdot \hbar \omega  \cdot \Xi_{ph} \left( \omega  \right) \cdot {\textrm d} \omega \\
        & = \frac{{k_B^2T}}{h}\int_0^{ + \infty } {\frac{{{x^2}{e^x}}}{{{{({e^x} - 1)}^2}}}}  \cdot \Xi_{ph} \left( {\frac{{x{k_B}T}}{\hbar }} \right) \cdot {\textrm d} x ,
        \end{align}
        where $x = \hbar \omega /{k_B}T$. This formula shows that thermal conductance is a weighted integration over the phonon transmission spectrum, and the weighting factor $x^2 e^x/(e^x-1)^2$ monotonically and rapidly decreases as the frequency of phonons increases. In fact, the weighting factor is less than $e^{-1}$ when $x>4$ and less than 1\% when $x>9$. As a result, one may safely say that thermal conductance is mainly determined by phonons with $\hbar\omega<10 k_BT$. The integration also implies that all phonons may contribute to thermal conduction, which is distinctly different to electronic conduction, where only a small portion of states, \textit{i.e.}, states around the Fermi energy, contribute to conduction properties.

        \textbf{Quantized thermal conductance.} Let us consider a ballistic transport system with only one phonon branch and at a very low temperature such that $k_BT\ll \hbar \omega_{\max}$, where $\omega_{\max}$ is the highest phonon frequency of the system. Intrinsic phonon transmission is 1 from $x=\hbar\omega/k_BT=0$ to $\hbar\omega_{\max}/k_BT\gg0$. Thus, thermal conductance of the system can be estimated to be
        \begin{align}
        {K_0}\left( T \right) &= \frac{{k_B^2T}}{h}\int_0^{\hbar {\omega _{\max }}/{k_B}T} {\frac{{{x^2}{e^x}}}{{{{({e^x} - 1)}^2}}}}  \cdot {\textrm d} x \cr
         & \approx \frac{{k_B^2T}}{h}\int_0^{ + \infty } {\frac{{{x^2}{e^x}}}{{{{({e^x} - 1)}^2}}}}  \cdot {\textrm d} x  = \frac{{{\pi ^2}k_B^2T}}{{3h}}.
        \end{align}
        This formalism shows that $K_0/T\approx 9.464310\times 10^{-4}~$nW/K$^2$. $K_0(T)$ is one quantized thermal conductance at temperature $T$ and proportional to $T$, representing the upper limit of thermal conductance that a single transport channel can have. Quantized phonon thermal transport was well studied and confirmed experimentally in 2000.\cite{Schwab_nat_2000} In fact, quantization of thermal conductance is universal, and not just limited to thermal transport of phonons.\cite{Pendry_1983,Maynard_PRB_1985,Angelescu_1998,Rego_PRL_1998,Blencowe_PRB_1999,yamamoto_prl_2004} Recently, quantized electrical thermal transport was also revealed in a single-atom gold junction at room temperature.\cite{Cui_science2017}

        \textbf{Transmission Spectrum.} For ideal crystalline structures, ballistic transmission equals to the number of transporting channels,\cite{WangJS_EPJ_2008,Xuyong_small_2014} which is
        \begin{align}
        \Xi_{ph} \left( \omega  \right) &= \sum\limits_\alpha ^{} {\left[ {\Theta \left( {\omega  - \omega _{\alpha ,\min }^{}} \right) - \Theta \left( {\omega  - \omega _{\alpha ,\max }^{}} \right)} \right]} \cr
        & = N\left( \omega  \right)
        \end{align}
        for a (quasi-)1D system, and\cite{Mingo_PRB_2003dispersion,Lundstrom_JAP_2010}
        \begin{align}
        \Xi_{ph} \left( \omega  \right) = {\sum\limits_{{{\alpha,\bf{k}}_ \bot }}^{} {} \Xi _\alpha ^{{{\bf{k}}_ \bot }}\left( \omega  \right)\left[ {\Theta \left( {\omega  - \omega _{\alpha ,\min }^{{{\bf{k}}_ \bot }}} \right) - \Theta \left( {\omega  - \omega _{\alpha ,\max }^{{{\bf{k}}_ \bot }}} \right)} \right]}
        \end{align}
        for a 2D or 3D system. Summation is carried out over phonon branches $\alpha$ and transverse $k$-vectors ${{{\bf{k}}_ \bot }}$.

        \textbf{Multi-probe systems.} From the derivation of thermal currents in previous sections, one can see that Equations~(\ref{eq:JL=-dHL}) and (\ref{eq:JL}) should be in the same form regardless of the number of thermal leads so long as there is no direct interaction between leads.
        The only thing that one needs to be concerned with is that the total self-energy should include self-energies of all probes now. The thermal current flowing from thermal probe $l$ to $l'$ is\cite{Buttiker_PRB1985,Dattabook,OuyangTao_PRB2017}
        \begin{align}
        {J_{Q;l \to l'}} = \frac{1}{{2\pi }}\int_0^{ + \infty } {\left[ {{f_{BE;l}}\left( \omega  \right) - {f_{BE;l'}}\left( \omega  \right)} \right]}  \cdot \hbar \omega  \cdot \Xi  _{ph;ll'}\left( \omega  \right) \cdot {\textrm d} \omega ,
        \end{align}
        where
        \begin{align}
        \Xi _{ph;ll'}\left( \omega  \right) = \Xi  _{ph;l'l}\left( \omega  \right)={\rm{Tr}}\left[ {{{\bf \Gamma} _{l'}}{{\bf G}^r}{{\bf \Gamma} _l}{{\bf G}^a}} \right]
        \end{align}
        is the transmission spectrum between $l$ and $l'$.

        \textbf{Crossover from ballistic to diffusive transport and the thermal conductivity.} Now we consider a transport system with length $L$ and suppose that the phonon mean free path (MFP) is $l_0$, writing the transmission spectrum as\cite{WangJian_APL2012,Mingo_PRB_2003GF,Dattabook}
        \begin{align}
        \Xi_{ph} \left( \omega  \right) = N\left( \omega  \right)\frac{{{l_0}}}{{{l_0} + L}}.
        \end{align}
        When the scattering region is much shorter than phonon MFP, \textit{i.e.}, $L\ll l_0$, the system is in the ballistic regime with
        \begin{align}
        \Xi_{ph} \left( \omega  \right) \approx N\left( \omega  \right).
        \end{align}
        When $L\gg l_0$, it is in the diffusive regime with
        \begin{align}
        \Xi_{ph} \left( \omega  \right) \approx N\left( \omega  \right){l_0}/L,
        \end{align}
        for which case the thermal conductance is
        \begin{align}
        K_{ph}\left( T \right)
        &=\frac{l_0}{L}\frac{1}{{2\pi }}\int_0^{ + \infty } {{{\left. {\frac{{\partial f_{BE}}}{{\partial T}}} \right|}_T} }  \cdot \hbar \omega  \cdot  N\left( \omega  \right) \cdot {\textrm d} \omega .
        \end{align}
        The number of phonon modes increases as the cross area gets larger, so we may estimate that $N(\omega)\propto A$. Under this estimation, the thermal conductivity becomes length($L$)-independent:
        \begin{align}
        \kappa_{ph} \left( T \right) &= K_{ph}\left( T \right)L/A\cr
        &=\frac{l_0}{{2\pi }}\int_0^{ + \infty } {{{\left. {\frac{{\partial f_{BE}}}{{\partial T}}} \right|}_T} }  \cdot \hbar \omega  \cdot \frac{ N\left( \omega  \right)}{A} \cdot {\textrm d} \omega .
        \end{align}
        This is consistent with the Fourier' Law for bulk materials, which states that thermal conductivity is a geometry-independent constant.

        \subsubsection{Phonon local density of states (LDOS)}
        Local density of states (LDOS) is very useful in investigating defects. Phonon LDOS can also be easily obtained using Green's functions, remembering that the oscillation spectrum $\omega_n$ is given by diagonalization of the dynamical matrix
        \begin{align}
        {\bf D} \left| {{u_n}} \right\rangle  = \omega _n^2\left| {{u_n}} \right\rangle ,
        \end{align}
        where $\left| {{u_n}} \right\rangle$ is the polarization vector of mode $n$.\cite{CallawayBook} For a non-interacting isolated system, the retarded Green's function can be written in the spectral representation as
        \begin{align}\label{eq:grSpectra}
        {{\bf G}^r}\left( \omega  \right) &= {\left[ {{{\left( {\omega  + {\textrm i}0^+ } \right)}^2}{\bf{I}} - {\bf D}} \right]^{ - 1}}
         = \sum\limits_n^{} {\frac{{\left| {{u_n}} \right\rangle \left\langle {{u_n}} \right|}}{{{{\left( {\omega  + {\textrm i}0^+ } \right)}^2} - \omega _n^2}}} \cr
        &  =\sum\limits_n^{} {\frac{{\left| {{u_n}} \right\rangle \langle {u_n}|}}{2\omega}    \left[ {P\left( {\frac{1}{{\omega  - {\omega _n}}}} \right) - {\rm{i}}\pi \delta \left( {\omega  - {\omega _n}} \right)} \right.} \cr
        &\left. { + P\left( {\frac{1}{{\omega  + {\omega _n}}}} \right) - {\rm{i}}\pi \delta \left( {\omega  + {\omega _n}} \right)} \right]
        \end{align}
        Here, $\bf{I}$ is an identity matrix, ``$P$'' means for the principal value, and the Plemelj formula is\cite{Wang_PRB2009}
        \begin{align}
        \mathop {\lim }\limits_{\eta  \to {0^ + }} \frac{1}{{\omega  + {\textrm i}\eta  \pm {\omega _n}}} = P\left( {\frac{1}{{\omega  \pm {\omega _n}}}} \right) - {\textrm i}\pi \delta \left( {\omega  \pm {\omega _n}} \right).
        \end{align}
        From Equation~(\ref{eq:grSpectra}), the imaginary part of ${\bf{G}}_{ii}^r$ is
        \begin{align}
        {\mathop{\rm Im}\nolimits} {\bf{G}}_{ii}^r\left( \omega  \right) =  - \frac{\pi }{{{\rm{2}}\omega }}\sum\limits_n^{} {|{u_{n;i}}{|^2}\left[ {\delta \left( {\omega  - {\omega _n}} \right)} \right.} \left. { + \delta \left( {\omega  + {\omega _n}} \right)} \right]
        \end{align}
        with ${u_{n;i}} = \left\langle i | {{u_n}} \right\rangle $ is the $i^{\textrm{th}}$ component of $\left| {{u_n}} \right\rangle$.
        If we focus on the range of $\omega\ge 0$, the term $\delta(\omega+\omega_n)$ vanishes. Then, according to the natural definitions of phonon LDOS and DOS, which are
        \begin{align}
        {\rho _i}\left( \omega  \right) &= \sum\limits_n^{} {|{u_{n;i}}{|^2}\delta \left( {\omega  - {\omega _n}} \right)},  \\
        \rho \left( \omega  \right) &= \sum\limits_n^{} {\delta \left( {\omega  - {\omega _n}} \right)} ,
        \end{align}
        respectively, we know that the phonon LDOS and DOS can be written in terms of retarded Green's function as
        \begin{align}
        {\rho _i}\left( \omega  \right) &=  - \frac{{2\omega }}{\pi }{\mathop{\rm Im}\nolimits} {\bf G}_{ii}^r\left( \omega  \right), \label{eq:ldos}\\
        \rho \left( \omega  \right)& =  - \frac{{2\omega }}{\pi }\sum\limits_i^{} {} {\mathop{\rm Im}\nolimits} \left\langle {{i}} \right|{{\bf G}^r}\left( \omega  \right)\left| {{i}} \right\rangle
         =  - \frac{{2\omega }}{\pi }{\rm Tr}{\mathop{\rm Im}\nolimits} {{\bf G}^r}\left( \omega  \right), \label{eq:dos}
        \end{align}
        respectively. Specifically, the phonon LDOS at $I^\textrm{th}$ atom is\cite{xuyongPhD}   
        \begin{align}
           {\rho _I}\left( \omega  \right) =- \frac{{2\omega }}{\pi }\sum_{\alpha=x,y,z}{\mathop{\rm Im}\nolimits} {\bf G}_{I\alpha, I\alpha}^r\left( \omega  \right).
        \end{align}

        According to Equations~(\ref{eq:ldos}) and (\ref{eq:dos}), phonon DOS and LDOS are given by the imaginary part of phonon retarded Green's function with a prefactor proportional to phonon frequency.

        \subsubsection{Acoustic sum rules}
        Due to rigid translational and rotational invariance, the force constants may satisfy
        \begin{align}
        \sum\limits_{I\alpha} ^{} \phi _{J\beta ,I\alpha } R_{I\alpha}^u =0, \quad u=1,\cdots,6,
        \end{align}
        where $\phi _{J\beta ,I\alpha }={\left. {\frac{{{\partial ^2}V}}{{\partial {R_{ J\beta}}\partial {R_{I\alpha }}}}} \right|_0}$ is a force constant, and $R^u_{I\alpha}$ describes a rigid motion of the lattice. For a finite-size (quasi-0D) system, there are 6 rigid motions including 3 translations and 3 rotations. Thus, for a 0D quantum dot, there should be 6 zero eigenmodes. For 1D periodic systems, 2 of the rigid rotations whose principal axis is perpendicular to the periodic direction break down due to the periodic boundary conditions. Consequently, 1D systems have 4 acoustic branches whose eigen frequencies are zero at the $\Gamma$ ($\mathbf{k}=0$) point. For 2D and 3D lattices, all rigid rotations are broken down, so only 3 acoustic branches exist. The acoustic sum rules place strong constraints on force constants and thus the dynamical matrix for all crystalline lattices. Therefore, the calculated force constants should be corrected to satisfy the acoustic sum rules to obtain the correct number of acoustic branches. Usually, one may subtract the nonzero sums from the background. To obtain accurate low-frequency transmission, force constants can be symmetrized via a Lagrange-multiplier symmetrization technique.\cite{Mingo_PRB2008}

        \subsubsection{Phonon-phonon interaction}

        In previous parts of this section, we introduced the basic phonon NEGF formalism for systems described by harmonic potentials. The anharmonic terms, \textit{e.g.}, the cubic or quartic terms of displacements, would cause phonon-phonon interactions. Without phonon-phonon interaction, thermal conductivity of a one-dimensional chain diverges, which is unphysical.\cite{Dhar_PRL_2008,XuYong_PRB2008} Therefore, phonon-phonon interaction is vital to the calculation of thermal conductivity of systems with sizes larger than their MFP. NEGF provides a natural and elegant, though sometimes complicated, way to incorporate phonon-phonon interactions.

        It is worth noting that there is an important assumption in NEGF transport theories that the (thermal) leads are noninteracting.\cite{Jauho_PRB_1994} In phonon cases, it means that phonons are noninteracting in the thermal leads and have various interactions in the central transport region. This assumption is reasonable because scattering results in finite lifetimes while eigen-states in the (thermal) leads have infinite lifetimes.

        A rigorous derivation about phonon-phonon interactions can be found in Ref.\citenum{WangJiansheng_PRE2007}. Here, we give only a brief description of the formalism. In the presence of phonon-phonon interactions, Equation (28) still holds and the effect of phonon-phonon interactions can be described by a self-energy term ${\bf \Sigma}_n$. Thus, the total self-energy can be expressed as\cite{WangJiansheng_PRE2007}
        \begin{align}
        {{\bf \Sigma} ^{\gamma}} = {\bf \Sigma} _L^{\gamma} + {\bf \Sigma} _R^{\gamma} + {\bf \Sigma} _n^{\gamma} \quad   ({\gamma}=r,a,<,>) .
        \end{align}
        In addition, Equation~(\ref{eq:GGG}) is also applicable, but with
        \begin{align}
        {\bf \Gamma}  = {{\bf \Gamma} _L} + {{\bf \Gamma} _R} + {{\bf \Gamma} _n},
        \end{align}
        where ${\bf \Gamma} _n={\textrm i}({\bf \Sigma}_n^r-{\bf \Sigma}_n^a)$. Substituting the above equations into Equation~(\ref{eq:JLw}), using the fact that $J_L=-J_R$ due to energy conservation, we have
        \begin{align}
        {J_L} = \frac{1}{2}\left( {{J_L} - {J_R}} \right) \approx \int_0^{ + \infty } {\frac{{{\textrm d}\omega }}{{2\pi }}}  \cdot \hbar \omega  \cdot {  \frac{\partial f_{BE}}{\partial T}} \cdot \Xi_{ph} \left( \omega  \right) \cdot \delta T.
        \end{align}
        Consequently, the thermal conductance in the linear response regime is
        \begin{align}
        K_{ph} =  \int_0^{ + \infty } {\frac{{{\textrm d}\omega }}{{2\pi }}}  \cdot \hbar \omega  \cdot \frac{\partial f_{BE}}{\partial T} \cdot \Xi_{ph} \left( \omega  \right) .
        \end{align}
        Here, the effective transmission function is defined as
        \begin{align}
        \Xi_{ph} \left( \omega  \right)
        &\equiv \frac{1}{4}{\rm{Tr}}\left\{ {{\bf G}^r\left( {2{{\bf \Gamma} _R} + {{\bf \Gamma} _n} + {\bf S}} \right){\bf G}^a{{\bf \Gamma} _L}} \right.\cr
        &\left.  + {\bf G}^r\left( {2{{\bf \Gamma} _L}+{{\bf \Gamma} _n} -{\bf  S}} \right){\bf G}^a{{\bf \Gamma} _R} \right\},
        \end{align}
        with
        \begin{align}
        {\bf S} \equiv \left. \left( {{f_{BE}} \frac{\delta{\bf \Gamma} _n}{\delta T} - {\textrm i} \frac{\delta {\bf \Sigma} _n^ <}{\delta T}  } \right) \right/\frac{\partial f_{BE}}{\partial T}.
        \end{align}

        \subsubsection{Summary}

\begin{figure}
          \centering
          \includegraphics[width=0.6\textwidth]{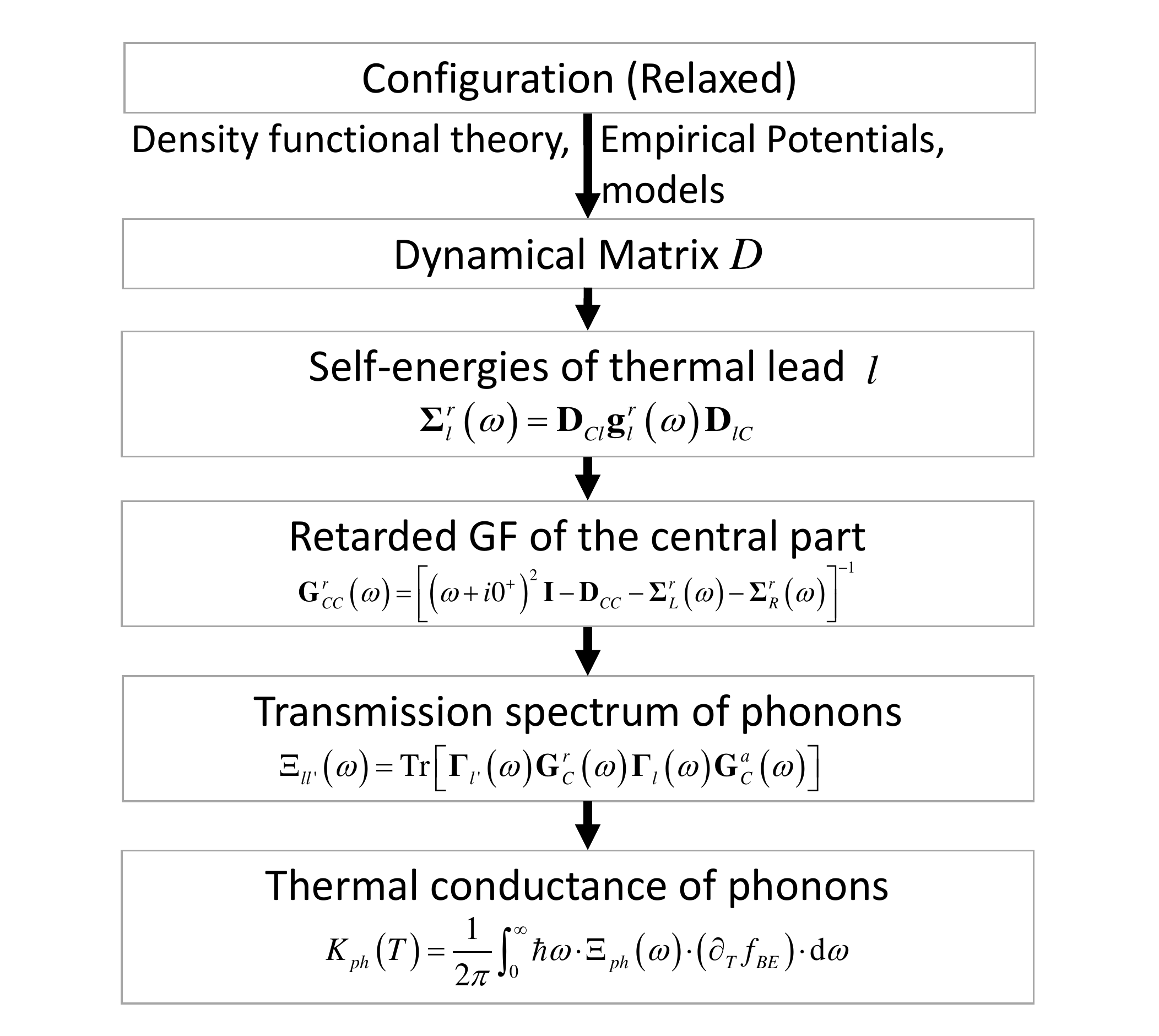}\\
          \caption{Work flow within the phonon NEGF framework in quasi-ballistic transport regime. 
          } \label{1-workflow}
    \end{figure}

        As a short summary for this section, we offer a schematic plot about how the phonon NEGF method works in Figure~\ref{1-workflow}. For practical applications, one needs to set up the configuration of the whole transport system. After structural relaxation, force constants can be obtained via empirical potentials,\cite{Xuyong_APL2009,cxb_PUCNT} density functional theory,\cite{TanZW_NanoLett_2011,Zou_nanoscale2015} or force constant models.\cite{SaitoBook} Then, the dynamical matrix of the system can be obtained immediately. With the knowledge of the dynamical matrix of ideal thermal leads, self-energies of individual thermal leads can be computed and introduced to the central region. To facilitate numerical calculation, several techniques can be applied to obtain the surface Green's functions rapidly.\cite{Sancho1985,xuyongPhD} Various scattering mechanisms, such as phonon-phonon interaction and electron-phonon interaction, can be also added to the central region. After obtaining all self-energies and the retarded Green's function of the central region, phonon transport properties, such as phonon transmission, thermal conductance, thermal current, phonon DOS, \textit{etc.}, can be calculated.

        Many exotic properties of thermal transport in nanoscale systems have been revealed by various methods.\cite{LiBaowen2017,XuBS_PDAP2016,HuJiuning_APL2010}
        In the next subsection, we shall focus on investigations carried out using the phonon NEGF method.

    \subsection{Phonon transport in nano devices}
    \subsubsection{Phonon transport through nanojunctions}
\begin{figure}
             \centering
          \includegraphics[width=0.6\textwidth]{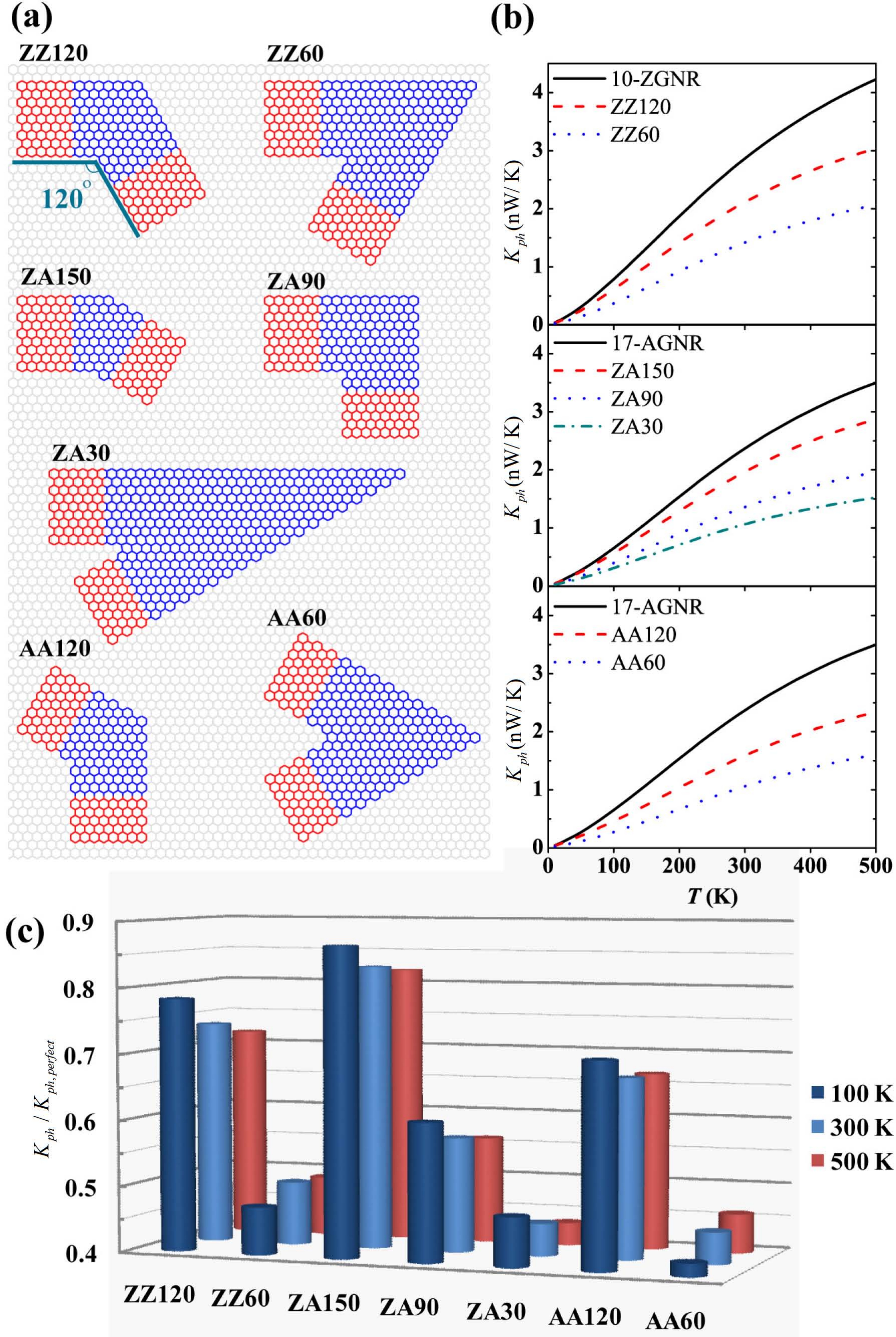}\\
          \caption{Thermal conduction in graphene junctions. (a) A graphene sheet can be cut into graphene homo-junctions with different connecting angles. (b) Thermal conductance $K_{ph}$ of graphene junctions with different connection angles. (c) Reduced thermal conductance, $K_{ph}/K_{ph,perfect}$ for different junctions at different temperatures.
          Figure adapted with permission from Ref.~\cite{Xuyong_prb_2010}. Copyrighted by the American Physical Society.}\label{fig:xyJunction}
    \end{figure}

        Carbon-based materials, such as graphene, are promising for electronic devices due to their ultra-high mobility and thermal conductivity.
        Having all components patterned in a graphene sheet, which has been experimentally demonstrated to have superior high thermal conductivity at room temperature, offers significant research potential.\cite{Balandin_NatMat_2011, Balandin_JPCM2012,PoP_MrsBull_2012,Xuyong_small_2014,RuanXiulin_NanoMicro2014}
        Although graphene itself lacks a band gap, graphene nanoribbons can be semiconducting or metallic, depending on their geometries. \cite{QiminYan_nanolett_2007} For example, metal-semiconductor junctions, $p$-$n$ junctions, heterojunctions, and field-effect transistors can be realized in patterned graphene junctions. It is thus necessary to understand how phonons transport in graphene-based nano structures for their future application in electronic devices. Xu $et~al.$ investigated various graphene junctions using the phonon NEGF method in the quasi-ballistic regime.\cite{Xuyong_prb_2010} They showed that phonon conduction favors junctions with a higher connection angle (Figure~\ref{fig:xyJunction}), which is opposite to the behavior of electrons. Classically, resistance should increase with the number of interfaces, however, the thermal resistance of double-interface junctions is only slightly higher than that of single-interface junctions. The main limiting factor is the narrowest part of the system. These findings are consistent with the picture of quasi-ballistic transport.

        Besides carbon-based circuits, molecular circuits are also very popular for the rapid development of molecular electronics.\cite{Cui_JCP2017}
        Kl\"ockner \emph{et~al.} studied the length dependence of phonon conduction in single-molecule junctions made of alkane chains of different number of methylene (CH$_2$) units and gold electrodes with connection via thiol or amine groups using the density-functional-theory-based phonon NEGF method.\cite{Klocker_PRB_2016} They demonstrated that the room-temperature thermal conductance is fairly independent of the length when the methylene units are more than five units. They also studied thermal conductance of C$_{60}$-based single-molecule junctions.\cite{Klockner_PRB2017} Due to weak metal-molecule coupling, phonon conduction is very weak ($\sim$O(10) pW/K at room temperature) in both Au-C$_{60}$-Au monomer junction and Au-C$_{60}$-C$_{60}$-Au dimer junction.

        Crucial in electronic devices,\cite{Jinhao_InSe2017} contacts are also vital in thermal conduction through nanojunctions. It was shown that the contact thermal resistance can contribute up to 50\% of the total measured thermal resistance in a 66-nm-diameter multiwalled CNT above 100 K.\cite{YangJK_small2011} Thermal contact resistance, which is defined as the temperature difference across an interface per unit heat flux,\cite{Chen_APL2009} can be used to represent heat conduction between interfaces. For example, the graphene-SiO$_2$ interface thermal resistance is measured to be $4.2\times 10^{-8}$,\cite{Freitag_NanoLett2009} $9.5\times10^{-8}$\cite{Jauregui_ECStran2010} ~K$\cdot$m$^2$/W through Raman spectroscopy for thermometry by different groups and 5.6$\times 10^{-9}\sim 1.2\times 10^{-8}$~K$\cdot$m$^2/$W through the $3\omega$ method by another group.\cite{Chen_APL2009}
        By engineering interfaces to lower thermal contact resistance, the heat generated at interfaces can be reduced."

\begin{figure}
           \centering
          \includegraphics[width=0.6\textwidth]{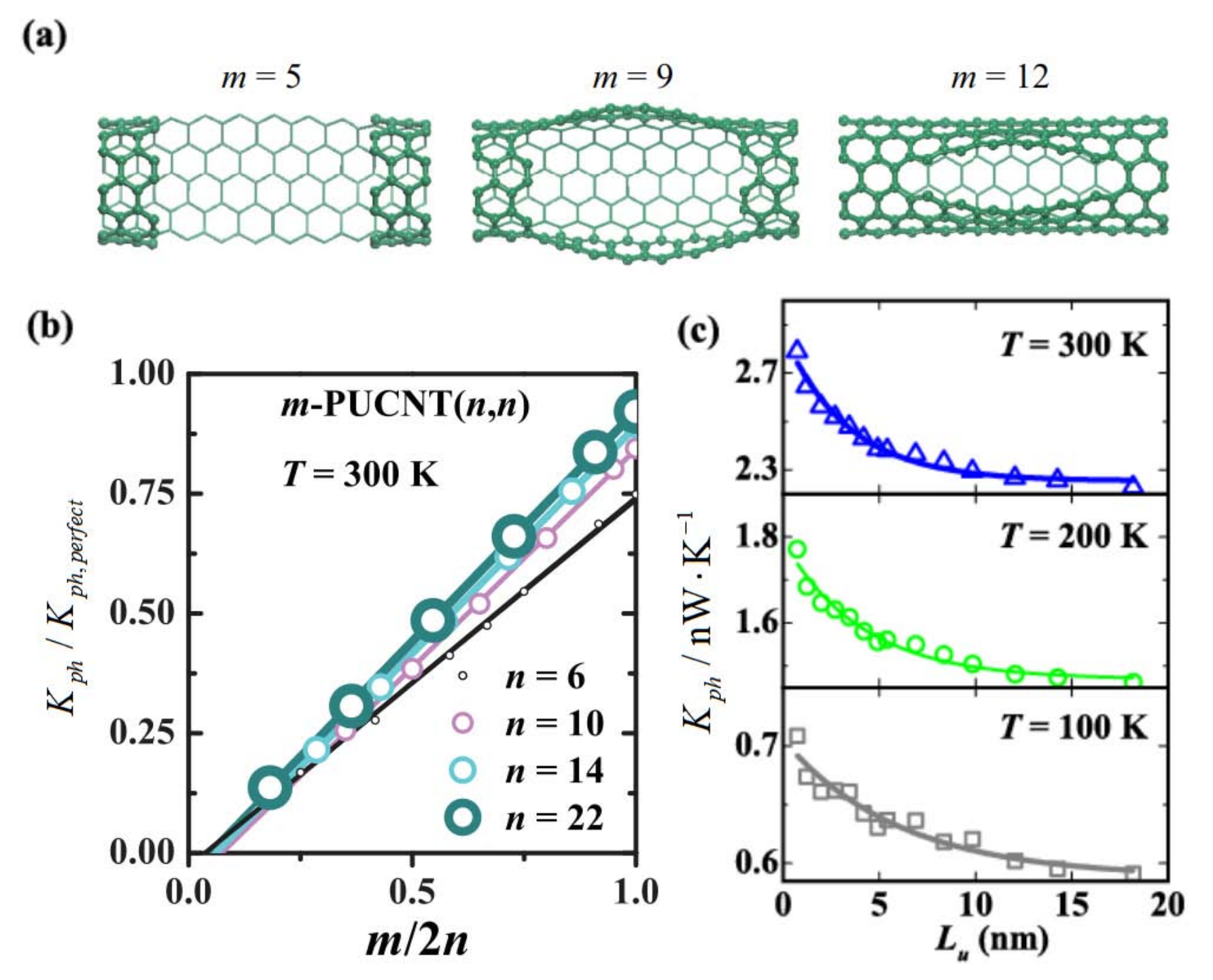}\\
          \caption{Interfacial thermal transport through carbon nanotube-nanoribbon interfaces: (a) Such interfaces are simulated using partially unzipped carbon nanotubes with $m$ lines of carbon dimers in the central area. (b) Scaled thermal conductance, which is the ratio of thermal conductance $K_{ph}$ of $m$-PUCNT($n$,$n$) to the thermal conductance of an ideal CNT($n$,$n$) at room temperature, as a function of scaled width $m/2n$. (c) Thermal conductance as a function of unzipped length $L_u$ at different temperatures.
          Figure adapted with permission from Ref.~\cite{cxb_PUCNT}. Copyrighted by the American Physical Society.}\label{fig:PUCNT}
    \end{figure}

        The interfaces between carbon nanotubes and nanoribbons would be common in carbon-based circuits. As revealed previously, thermal conductance is restricted by the narrowest part in graphene nanojunctions. One may wonder if this \emph{bottleneck effect} is applicable in other carbon materials and even other non-carbon nano materials. In the interface between a carbon nanotube (CNT) and a curved graphene nanoribbon in a partially unzipped carbon nanotube (PUCNT) [Figure~\ref{fig:PUCNT}(a)],\cite{cxb_PUCNT} Chen $et~al.$ showed that the thermal conductance is roughly proportional to the reduced width of the nanoribbon region. In addition, the linearity of the thermal conductance to the reduced width of the nanoribbon implies that the main restricting factor for phonon conduction is the width of the central part, which is represented by $m$---the number of carbon dimer lines in the center.
        Because the unzipped part always contains less phonon modes than the other parts, the bottleneck effect works. Similar linearity is also found in silicon nanowires.\cite{Mingo_PRB2007} Besides intrinsic restriction brought by the narrow part, the length dependence of thermal conductance for a PUCNT shows an exponential variation and approaches to nonzero values. [Figure~\ref{fig:PUCNT}(b)] The exponential decay behavior can be attributed to scattered unmatched modes, and the nonzero limit can be attributed to transmitted matched modes. These arguments were satisfactorily supported by a one-dimensional atomic chain model. The linearity and exponential decay to nonzero values are defining features in quantum thermal transport.\cite{cxb_PUCNT}

        Interfaces in silicon-based circuits are a concern for current integrated circuits. Miao $et~al.$ extended the B\"uttiker probe approach previously used for electron transport problems to the phonon NEGF method to empirically and efficiently include various scattering mechanisms, such as impurity, boundary, and Umklapp scattering.\cite{Fisher_APL2016} Various scattering mechanisms were described by different empirical forms of the scattering rates $\tau^{-1}$ and were introduced through diagonal effective self-energies of one B\"uttiker probe as
        \begin{align}
        {\bf{\Sigma}}_{BP}^r=-2{\textrm i} \frac{\omega}{\tau} \bf{I},
        \end{align}
        where $\bf{I}$ is an identity matrix.
        Sadasivam $et~al.$ demonstrated later that inelastic process also plays an important role in epitaxial CoSi$_2$-Si interfaces.\cite{Fisher_PRB2017} By incorporating phonon-phonon and electron-phonon couplings in both the metal and interfaces, they obtained results consistent with experiments.

        Grain boundaries (GBs) are one special kind of interface. They occur at the boundary of two grains of the same material and are ubiquitous in chemical vapor deposition (CVD)-synthesized samples.\cite{ZouXL_AccChem2015,NanoLett2013_XZou,ZouXL_small2015} The zero-degree grain boundaries are usually referred to as extended line defects. The pentagon-heptagon extended defect has a stronger impact on thermal condutance in armchair-oriented graphene nanoribbons (AGNRs) than in zigzag-oriented GNRs (ZGNRs).\cite{ChenKeqiu_Carbon2013}
        The thermal conductance can be tuned through the orientation and bond configuration of the extended defect by over 50\% at room temperature.
        Such a decrease can be attributed to the modification of phonon dispersion and/or to the tailoring of the strength of defect scattering.\cite{Huaqing_PRB_2013} When edges are reconstructed with pentagon-heptagon edges, extremely narrow ZGNRs may have up to 75\% of their original thermal conductance suppressed.\cite{LanJinhua_JPDAP2014} Further investigation on CNTs with extended line defects using the phonon unfolding method shows that the phonon spectra of defected CNTs are split with obvious gaps opening, leading to lower phonon transmission.\cite{Huaqing_JPCM2015}

    \subsubsection{Anisotropic thermal conduction} 

\begin{figure}
        \centering
          \includegraphics[width=0.6\textwidth]{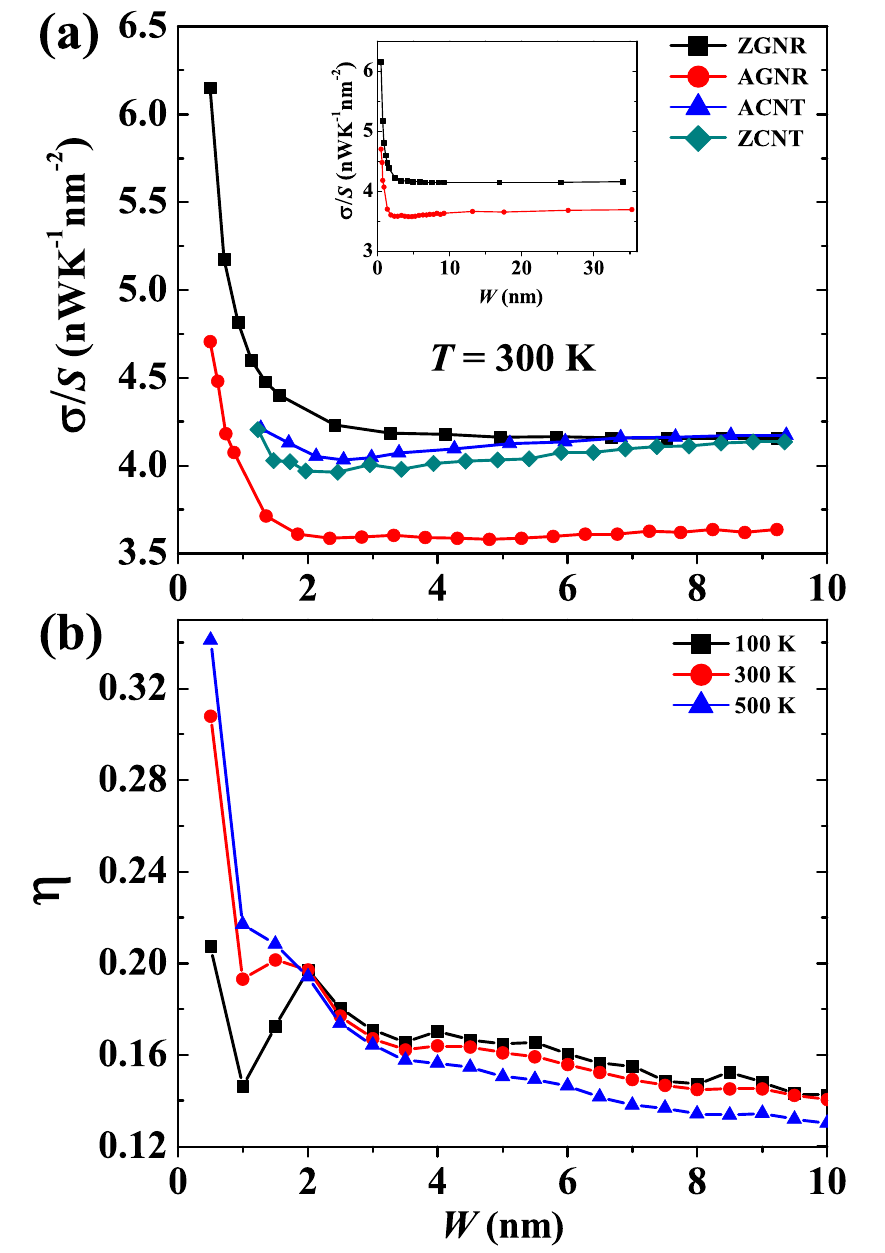}\\
          \caption{Anisotropy of thermal conductance in graphene. (a) Room-temperature thermal conductance of zigzag/armchair graphene nanoribbons (ZGNRs/AGNRs) and carbon nanotubes (ZCNTs/ACNTs) scaled by cross-sectional areas ($S$) as a function of width. (b) Anisotropy factor, $\eta=[(\sigma/S)_{\textrm {ZGNR}}-(\sigma/S)_{\textrm {AGNR}}]-1$ , as a function of width. Reprinted with permission.\cite{Xuyong_APL2009} Copyright 2009, AIP Publishing.
         }\label{fig:anisotropy}
    \end{figure}

        Anisotropic materials are those whose thermal conductance/conductivity are different along different directions. For example, the room-temperature thermal conductance of $\langle110\rangle$ pristine silicon nanowires (SiNWs) is higher than that of $\langle100\rangle$ and $\langle111\rangle$ SiNWs by 50\% and 70\%, respectively.\cite{Markussen_nanoLett2008} This anisotropy originates from phonon band structures.
        For graphene nanoribbons, whose electronic properties are extremely sensitive to device geometry, thermal conduction also has large anisotropy as revealed by phonon NEGF calculation.\cite{Xuyong_APL2009} Xu $et~al.$ showed that the room-temperature thermal conductance of ZGNRs is up to 30\% higher than that of AGNRs. In addition, the anisotropy decreases as the width of nanoribbon increases and is expected to disappear when the width of GNRs exceeds 100 nm. By contrast, carbon nanotubes, which can be deemed as rolled from a graphene nanoribbon, have negligible anisotropy[Figure~\ref{fig:anisotropy}], implying that the large anisotropy in thermal conductance of graphene nanoribbons can be attributed to different boundary conditions of AGNRs and ZGNRs. Furthermore, Tan $et~al.$ attributed this anisotropy to band dispersion.\cite{TanZW_NanoLett_2011}
        Anisotropic thermal conductance has also been found in monolayer boron nitride,\cite{OuyangTao_Nanotech_2010,ChenYuanping_Nanotech2016} graphyne nanoribbons,\cite{OuyangTao_PRB_2012,OuyangTao_Nanotech2014} graphane nanoribbons,\cite{LiDengfeng_APL_2014} curved graphene nanoribbons,\cite{LiDengfeng2013}
        and monolayer black phosphorus.\cite{ZhangYongwei_JPCC2014,WuJunqiao_NatComm2015,ShiLi2016} Yet, anisotropic thermal conduction seems to be absent in stanene.\cite{ZhangGang_PRB2016}

   \subsubsection{Point Defects}
       Scattering from point defects, such as vacancies, interstitial atoms, and impurities, may also have a remarkable impact on phonon transport properties besides bulk thermal properties.\cite{ZhuBangfen_PRB_2005}
       When the concentration of point defects is low ($<10\%$\cite{Mingo_PRL_2008isotope}), multiple-scattering-induced interference effects can be ignored in the diffusive regime, and the total transmission is given by the Cascade scattering model:\cite{Mingo_PRL_2008isotope,Jauho_PRL_2009,LiDengfeng_APL_2014}
        \begin{equation} \label{eq:Tscat}
        \frac{1}{{{ \bar \Xi}_n}} = \frac{n}{{{{\bar \Xi}_1}}} - \frac{{n - 1}}{{{{\bar \Xi}_{0}}}} ,
        \end{equation}
        where $n$ is the number of vacancies, ${{\bar \Xi}_n}({{\bar \Xi}_1})$ is the configuration-averaged transmission with the presence of $n$(1) vacancies, and ${{\bar \Xi}_0}$ is the ballistic transmission without defects. When the sizes of the system are large enough such that $n\to\infty$, transmission scales as $\propto1/n$, implying that scattering from point defects cannot be ignored in large-size samples even if the concentration of defects is low.

        Significant reduction of thermal conductance by hydrogen vacancies is observed in graphane nanoribbons.\cite{LiDengfeng_APL_2014} Isotope impurities, which are usually point defects for phonons, may also lead to the reduction of thermal conductance. Savi\'c \emph{et~al.} revealed a significant reduction of thermal conductance by isotope impurities in both carbon and boron-nitride nanotubes using $ab~initio$ calculation of force constants.\cite{Mingo_PRL_2008isotope} For example, a 2.6 $\mu$m-long carbon nanotube with 10.7\% $^{14}$C isotopes retains only 20\% of its pristine thermal conductance.
        However, when the temperature approaches 0 K, the thermal conductance approaches the universal quantum value $NK_0$ with $N$ acoustic phonon branches, even with the presence of local point defects, such as the SW defect and a single vacancy defect.\cite{yamamoto_PRL2006,ChenKeqiu_PRB2010,PengXF_Carbon2014}

    \subsubsection{Strain Effects}
         Strain, which is unavoidable in real applications, is widely used to tune both thermal\cite{ZhangGang_MechMater2015,Lindsay_PRB2014,GongXingao_APL2009} and electronic\cite{SiChen_Nano2016,Menglei2015,zhoumei,zhouMei2015,Jinhao_strain2017} properties of low-dimensional materials.

        Jin $et~al.$ considered the strain effects on force constants by elastic theory and found that thermal conductance of graphene nanoribbons under uniaxial stretching strain could be enhanced by up to 17\% and 36\% for 5nm-wide ZGNRs and AGNRs, respectively.\cite{JinGuojun_EPL2011} Thermal conductance of tensile-strained ZGNRs is higher than unstrained ones because of the modulation of $\pi$-orbital overlap integral.\cite{PengXF_Carbon2016}

        When CNTs contact with substrates, radial compression occurs. High radial compression has been known to induce metal-insulator transitions.\cite{LuJunqiang_PRL2003} Zhu $et~al.$ investigated the effect of radial strain on thermal conductance of bended carbon nanotubes and revealed a robust linear dependence of thermal conductance on radial strain in CNTs.\cite{ZhuHongqin_NJP2012}

 \subsection{Generalized phonon NEGF method and topological phonon devices}
    The topological states of quantum matters have revolutionized the research of condensed matter physics and material science. Topological states, such as the quantum (anomalous/spin) Hall [Q(A/S)H] state, can provide dissipationless conduction channels at edges and are thus useful for power-consuming electronics, thermoelectrics, spintronics, and topological computations. At the same time, large efforts have been devoted to manipulating heat flow because the ability to control heat can potentially lead to the development of phononic devices for thermal barrier coating, thermal isolation, and even thermal based information processing.\cite{LiBaowen_RMP_2012} Recently topological concepts have been applied to phonon systems, leading to an emerging field of ``\textit{topological phononics}''. In the following, we give a brief introduction to the application of the NEGF method to topological phonon systems.

    Phonon topology is closely related to the time reversal symmetry (TRS) of phonons. The integration of Berry curvature over the 2D Brillouin zone defines a topological invariant---Chern number---which is odd under TRS operation. Thus, TRS must be broken in a topological phonon system with nonzero Chern number. \cite{Xiao2010} However, in previous treatments of phonons, TRS is explicitly assumed. The TRS-broken effects of phonons can be incorporated into the following lattice Lagrangian \cite{Liu2017a}
    \begin{equation}
    L = \frac{1}{2}\dot{u}_i\dot{u}_i - \frac{1}{2}D_{ij}u_iu_j + \eta_{ij}\dot{u}_iu_j,
    \end{equation}
    in which the Einstein sum rule has been used. The first two terms are the ordinary kinetic energy term and harmonic potential energy terms, respectively, while, the last term is the TRS breaking term, which couples displacements and velocities. The dynamic equation of motion is governed by a phonon Schr\"odinger-like equation: $\mathbf{H_k} \psi_{\mathbf{k}} = \omega_{\mathbf{k}} \psi_{\mathbf{k}}$ \cite{Liu2017a} where
    \begin{equation}
    \mathbf{H_k} = \left(
    \begin{array}{cc}
    \mathbf{0} & \mathrm{i}\mathbf{D}^{1/2}_{\mathbf{k}} \\
    \mathrm{i}\mathbf{D}^{1/2}_{\mathbf{k}} & -2\mathrm{i}\bm{\eta}_{\mathbf{k}}
    \end{array}
    \right), \quad
    \psi_{\mathbf{k}} = C \left(
    \begin{array}{c}
    \mathbf{D}^{1/2}_{\mathbf{k}} \mathbf{u_k} \\
    \dot{\mathbf{u}}_{\mathbf{k}}
    \end{array}
    \right).
    \end{equation}
    In this equation, $\omega_{\mathbf{k}}$ represents the phonon dispersion relations and can take both positive and negative values, satisfying $\omega_{\mathbf{k}} = -\omega_{-\mathbf{k}}$. $C$ is a normalization constant. Similar equations with different definitions of wavefunctions have been proposed. \cite{ZhangLifa_PRL2010, susstrunk2016} When TRS is present, \textit{i.e.}, $\bm{\eta}_{\mathbf{k}} =\bm{0}$, the phonon Schr\"odinger-like equation reduces to $\mathbf{D_k u_k} = \omega^2_{\mathbf{k}} \mathbf{u_k}$.

    For phonon transport, the standard NEGF formalism introduced earlier in this review works only for systems with TRS (\textit{i.e.}, $\bm{\eta}_\mathbf{k}=\bm{0}$). Based on the phonon Schr\"odinger-like equation of motion, we can derive a generalized NEGF approach, which applies to TRS-broken cases as will be illustrated below. Intuitively, one can define the retarded (advanced) Green's function as
    \begin{align}
    {\bf G}^{r,a }(\omega) = [(\omega\pm {\textrm i}0^+){\bf I} - {\bf H}]^{-1} ,
    \end{align}
    resembling Green's functions of electrons. These Green's functions are different to the traditional phonon Green's functions defined in Equation~(\ref{eq:Gra}). Here, ${\bf H}$ contains the square root of the dynamical matrix, ${\bf D}^{1/2}$, and because taking a square root may lead to long-ranged intersites' couplings, we cannot decouple the left and right leads, which makes the calculation much more complex. Thus, we cannot decouple the left and right leads, which makes the calculation much more complex. To avoid this problem, we can change the phonon equation by adopting a ``local'' phonon wavefunction $\tilde{{\bf y}} = (\mathbf{u}^T, \dot{\mathbf{u}}^T)^T$ ($T$ stands for matrix transpose), modifying the equation to ${\bf Q}\tilde{{\bf y}} = \omega {\bf R}^{-1} \tilde{{\bf y}}$,\cite{Liu2017a} where
    \begin{equation}
    {\bf Q} = \left(
    \begin{array}{cc}
    {\bf D} & {\bf 0} \\
    {\bf 0} & {\bf I}
    \end{array}
    \right), {\bf R} = \left(
    \begin{array}{cc}
    {\bf 0}   & {\textrm i}{\bf I} \\
    -{\textrm i}{\bf I} & -2{\textrm i}\bm{\eta}
    \end{array}
    \right).
    \end{equation}
    Thus, we can define a new Green's function without taking square root of the dynamical matrix:
    \begin{align}
    {\bf G}^{r,a}(\omega) = [(\omega+{\textrm i}0^+){\bf R}^{-1} - {\bf Q}]^{-1}.
    \end{align}
    It should be noted that the new Green's functions are defined in the extended coordinates-velocities space and the interatomic coupling is
    \begin{equation}
    {\bf U} = \left(
    \begin{array}{cc}
    {\bf V} & {\bf 0} \\
    {\bf 0} & {\bf 0}
    \end{array}
    \right),
    \end{equation}
    where ${\bf V}$ refers to the ordinary coupling (\textit{i.e.}, ${\bf D}_{CL}$, ${\bf D}_{CR}$, \textit{etc.}). The self-energy from the thermal lead $l$ is determined by
    \begin{align}
    {\bf \Sigma}^{r,a}_l = {\bf U}_{Cl}{\bf g}^{r,a}_l{\bf U}_{l C} \quad (l=L, R).
    \end{align}
    Finally, the phonon transmission function is given in the same form with Equation~(\ref{eq:trans}).

    As an example, topological phonons can be realized in a honeycomb lattice where the longitudinal optical (LO) and longitudinal acoustic (LA) phonon branches linearly cross and form a pair of Dirac cones at BZ corners $K$ and $K^\prime$.\cite{Liu2017a} The phonon modes near the Dirac points can be described by an effective Hamiltonian:
    \begin{equation}
    H_0 = v_D (k_y\tau_z\sigma_x - k_x\sigma_y),
    \end{equation}
    where $v_D$ is the group velocity, $\sigma$ and $\tau$ are Pauli matrices with $\sigma_z=\pm1$, and $\tau_z=\pm1$ indicating the $A(B)$ sublattice and $K(K^\prime)$ valley, respectively. To break the crystalline symmetry, or TRS, one may introduce mass terms into the effective Hamiltonian. Generally, the Dirac mass terms should \textit{anti}commute with $H_0$, implying four types of independent mass terms $\sigma_z\tau_z$, $\sigma_z$, $\sigma_x\tau_x$, and $\sigma_x\tau_y$.\cite{Liu2017c} $\sigma_z\tau_z$ (the Haldane term) breaks TRS and can be realized either extrinsically by applying an external magnetic/Coriolis field or intrinsically by Raman spin-lattice interactions in magnetic lattices. \cite{Liu2017a} $\sigma_z$ (the Semenoff term) breaks inversion symmetry (IS) and can be realized using different atomic masses for $A$, $B$ sublattices. The $\sigma_x\tau_x$ and $\sigma_x\tau_y$ terms mix $K$ and $K^\prime$ valleys and can be realized by Kekul\'e distortion. \cite{Liu2017c} Generally, the mass terms can be expressed as
    \begin{equation}
    \begin{split}
    H^\prime &= H_T + H_I + H_K \\
             &= m_T\sigma_z\tau_z + m_I\sigma_z + m_K\sigma_x\tau_n,
    \end{split}
    \end{equation}
    where $\tau_n=\tau_x\cos\theta + \tau_y\sin\theta$.

\begin{figure*}
      \centering
    \includegraphics[width=\linewidth]{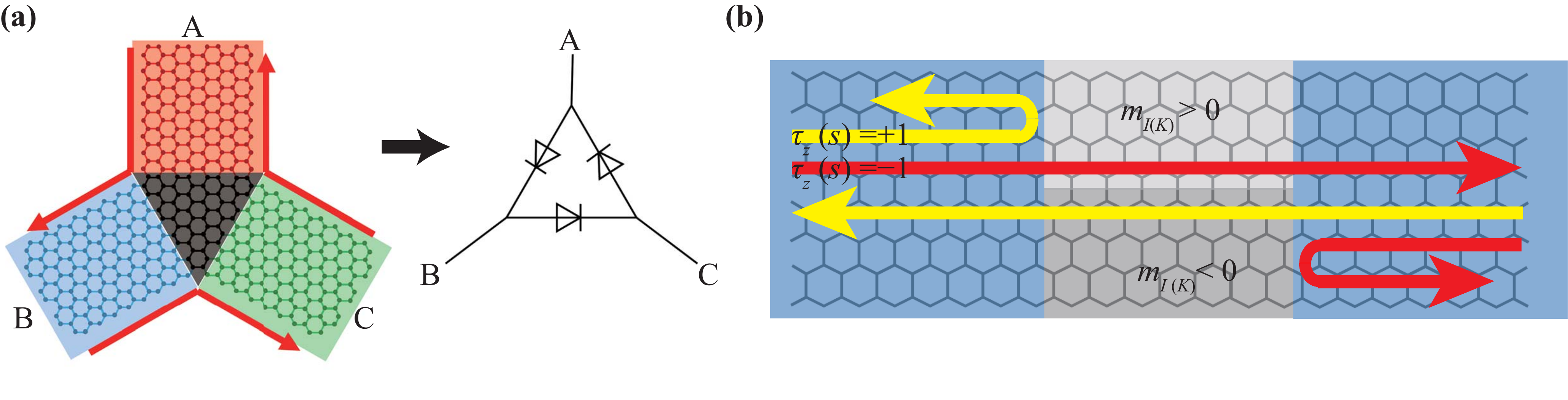}
    \caption{\label{TopoDevices}
    Schematic models of topological phononic devices including perfect phonon diode and phonon valley- (pseudospin-) filter. (a) Schematic diagram of QAH-like one-way edge states of a honeycomb strip with three terminals. Perfect diode effect exists between any two of the terminals A, B, and C. (b) Schematic diagram of phonon current through a topological region which is composed by two parts in parallel with $m_{I(K)}>0$ and $m_{I(K)}<0$, respectively. Phonon modes labelled by certain valley (pseudospin) index [\textit{i.e.}, $\tau_z (s)$] are allowed to travel unidirectionally due to the valley- (pseudospin-) momentum locking.  }
    \end{figure*}

    Topological phonons have various promising applications for phonon devices. One possibility is based on the scattering-free one-way boundary states, \textit{i.e.}, an ideal phonon diode can be achieved in a multiterminal system [Figure \ref{TopoDevices}(a)]. Another possibility is based on the valley (pseudospin) degree of freedom of phonon. A pure valley (pseudospin) current of phonons can be realized in a topological domain boundary where $m_{I(K)}$ changes sign [Figure \ref{TopoDevices}(b)].

   Several topological phononic models have been realized experimentally in recent studies. In 2015, Nash \textit{et al.} built a honeycomb lattice composed by gyroscopes and springs. \cite{Nash2015} The spinning of gyroscopes breaks TRS, which mimics Coriolis/magnetic field or, equivalently, the Haldane model $H_T$ as mentioned above. The resulting lattice is a phononic analogue of QAH effect. In the same year S{\"u}sstrunk \textit{et al.} realized the mechanical QSH effect in a lattice of pendula. \cite{Susstrunk2015} They also experimentally observed the helical edge modes of the mechanical wave. In 2016, He \textit{et al.} successfully made an acoustic topological insulator by connecting two types of phononic ``graphene'' consisting of strainless-steel rods in air. \cite{he2016} In the same year, a TRS-invariant Floquet topological insulator for sound was achieved by Peng \textit{et al}. \cite{peng2016} Also, topological modes were observed in Maxwell frames. \cite{Chen2014, Paulose2015a, Paulose2015b, Chen2016, Meeussen2016}

\section{Caloritronic devices}
    \subsection{Thermoelectricity}
        Thermoelectricity is the interconversion between heat and electrical currents. It can be used for power generation and cooling. The study of thermoelectric effects has attracted much research interest because of the stability, reliability, and scalability of thermoelectric devices. Low-dimensional materials are promising for thermoelectric applications due to their high carrier mobility, tunable band gaps, and the quantum effects of confinements. The NEGF methods are useful for studying thermoelectric effects in nanostructures. In this section, we first give a basic formalism about thermoelectricity, on which a dimensionless figure of merit $ZT$ can be naturally defined. Second, we introduce the charge/spin figure of merit $Z_{c/s}T$. Finally, we discuss some investigations of thermoelectric, spin caloritronic, and valley caloritronic devices.

        \subsubsection{Linear Response Theory}
            Similar to the formalisms in previous sections, electronic current and electronic thermal current flowing from spin-degenerate lead $L$ to $R$ can be written as\cite{Dattabook}
            \begin{align}
                    I &= \frac{{2e}}  {h}\int { \mathcal{T} \left( E \right)\left( { f_{FD;L}  - f_{FD;R}} \right){\textrm d} E },\\
                    I_Q &= \frac{{2}}  {h}\int {(E-\mu) \mathcal{T}\left( E \right)\left( f_{FD;L} - f_{FD;R} \right){\textrm d} E },
            \end{align}
            where $f_{FD;l}(E)=1/[e^{(E-\mu_l)/k_BT_l}+1]$ is the Fermi-Dirac distribution of the lead $l$, $e<0$ is the elementary charge, and $\mathcal{T}(E)$ is the electronic transmission spectrum. The factor of 2 is due to spin degeneracy.
            Under linear response conditions, both $I$ and $I_Q$ can be linearly expanded as\cite{MahanBook}
            \begin{align}
            I  &\approx\frac{{2e}}
            {h}\int {T\left( E \right)\left[ { - (f_{FD})_E' \Delta \mu  - (f_{FD})_E' \left( {\frac{{E - \mu }}
            {T}} \right)\Delta T} \right]{\textrm d} E }  \hfill  \notag \cr
               & \equiv eL_0 \Delta \mu  + \frac{e}
            {T}L_1 \Delta T \hfill  \notag \\
               &= e^2 L_0 \Delta V + \frac{e} {T}L_1 \Delta T, \\
            I_Q  &= \frac{2}
            {h}\int {T\left( E \right)\left( {E - \mu } \right)\left( f_{FD;L}  - f_{FD;R}  \right){\textrm d} E }  \cr
            &= eL_1 \Delta V  + \frac{1}
            {T}L_2 \Delta T, \label{the:eq:IQ}
            \end{align}
            with $\Delta\mu=e\Delta V$, $\Delta V=V_L-V_R$, $\Delta T=T_L-T_R$, and
            \begin{equation}\label{eq:Ln}
            L_n  \equiv \frac{2}
            {h}\int {\mathcal{T}\left( E \right)\left( {E - \mu } \right)^n \left( { - \frac{\partial f_{FD}}{\partial E} } \right){\textrm d} E }.
            \end{equation}
            The transmission function for electrons is obtained through the Caroli formula
            \begin{equation}
            \mathcal{T}(E) = {\rm{Tr}}\left[ {{{\bf{\Gamma }}_L}{{\bf{G}}^r}{{\bf{\Gamma }}_R}{{\bf{G}}^a}} \right].
            \end{equation}
            Here, ${\bf{\Gamma }}_{L/R}$ and ${{\bf{G}}^{r,a}}$ are the electronic version of bandwidth function and Green's functions, respectively. More details can be found in Refs.~\citenum{Jauho_Book} and \citenum{Dattabook}.

            Now, we can write these response equations into a matrix form as
            \begin{align}\label{eq:JandJQ}
            \left( \begin{gathered}
              I \hfill \\
              I_Q  \hfill \\
            \end{gathered}  \right) = \left( {\begin{array}{*{20}c}
               {eL_0 } & {eL_1/T}  \\
               {L_1 } & {L_2/T }  \\
             \end{array} } \right)\left( \begin{gathered}
              e\Delta V \hfill \\
              \Delta T \hfill \\
            \end{gathered}  \right).
            \end{align}

        \subsubsection{Thermoelectric coefficients \label{sec:ZT}}

            Thermoelectric effects include the Seebeck effect, Peltier effect, and Thomson effect. The transport coefficients, such as the electrical conductance $G$, Seebeck coefficient $S$, and the electronic thermal conductance $K_e$, are relevant for the calculation of thermoelectric energy conversion efficiency. Thus, in the following we shall represent these coefficients in terms of $L_n$, which is defined in Equation~(\ref{eq:Ln}).

            To begin with, from Ohm's law,
            \begin{align}
            I = G\Delta V,
            \end{align}
            we know that
            \begin{align}
              G = e^2 L_0
            \end{align}
            by making a comparison to Equation~(\ref{eq:JandJQ}).

            Second, the Seebeck coefficient, which is also called the ``thermopower'',\cite{MahanBook} measures the open-circuit thermoelectric voltage generated by one unit temperature gradient as
            \begin{align}\label{eq:s}
            S =  - {\left. {\frac{{\Delta V}}{{\Delta T}}} \right|_{I = 0}}.
            \end{align}
            The magnitude of the Seebeck coefficient for a good thermoelectric material, \textit{e.g.}, Bi$_2$Se$_3$, amounts to about $10^2$ $\mu$V/K. Usually, a larger band gap favors a higher optimized Seebeck coefficient. Let $I=0$ in Equation~(\ref{eq:JandJQ}), and we get the Seebeck coefficient following the definition of Equation (\ref{eq:s}) as
            \begin{align}
            S = \frac{e} {T}L_1 /e^2 L_0  = -\frac{{L_1 }}{{|e|TL_0 }}.
            \end{align}

            Finally, the electronic thermal conductance $K_e$ is defined as the thermal conductance of electrons when there is no electric current. Thus, by combining $I=0$ and $I_Q=-K_e \Delta T$ in Equation~(\ref{eq:JandJQ}), we get
            \begin{align}
                 K_e  = \frac{1}{T}L_2  - GS^2 T = \frac{1}{T}\left( {L_2  - \frac{{L_1 ^2 }}
            {{L_0 }}} \right).
            \end{align}

            Replacing $L_n$ in Equation~(\ref{eq:JandJQ}) by $G$, $S$, and $K_e$, we have
            \begin{align}
            I & = G\Delta V + GS\Delta T, \label{eq:I=GV+GST} \\
            {I_Q} &= \left( {{K _e} + G{S^2}T} \right)\Delta T + GST\Delta V\cr
                  &= {K _e}\Delta T + STI, \label{eq:IQ}
            \end{align}
            or in the matrix form as
            \begin{align}\label{eq:IIQ}
            \left( \begin{array}{l}
            I\\
            {I_Q}
            \end{array} \right) = \left( {\begin{array}{*{20}{c}}
            G&{GS}\\
            {GST}&{{K _e} + G{S^2}T}
            \end{array}} \right)\left( \begin{array}{l}
            \Delta V\\
            \Delta T
            \end{array} \right).
            \end{align}

             Also, let us look at the situation $\Delta T=0$ in Equation~(\ref{eq:IQ}), which reads
             \begin{align}
             {I_Q} = STI.
             \end{align}
              This equation shows that there is a thermal current $I_Q$ which accompanies the charge current $I$ without the presence of temperature gradient, and the proportional coefficient is
             \begin{align}
             \Pi \equiv I_Q/I=ST.
             \end{align}
             This is the Peltier effect, and $\Pi$ is the Peltier coefficient.

             As a short summary, we list the formulas for $G$, $S$, and $K_e$ in the following
            \begin{align}
              G& = e^2 L_0 , \hfill  \label{G}\\
              S &= \frac{e} {T}L_1 /e^2 L_0  = -\frac{{L_1 }}{{|e|TL_0 }}, \hfill   \label{S}\\
              K_e & = \frac{1}{T}\left( {L_2  - \frac{{L_1 ^2 }}{{L_0 }}} \right) \label{k},
            \end{align}
            with $L_n$ defined in Equation~(\ref{eq:Ln}).

\begin{figure}
                    \centering
              \includegraphics[width=0.6\textwidth]{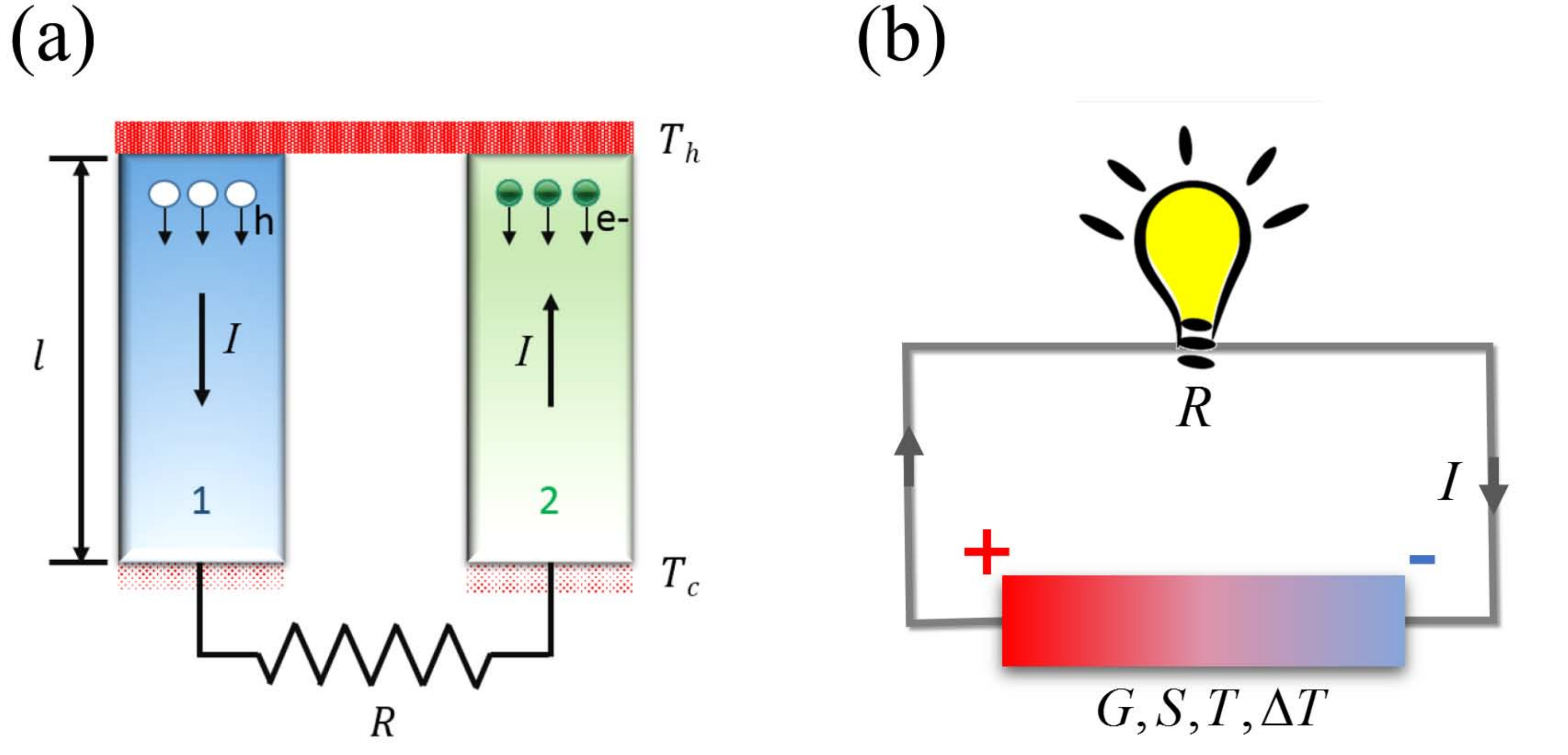}\\
              \caption{Thermoelectric devices. (a) One thermoelectric couple consists of $n$-type and $p$-type materials as two legs. When one end of the thermoelectric couple is heated, $T_h>T_c$, electric current flows in the circuit. A thermoelectric generator can be made by connecting thermoelectric couples electrically in series and thermally in parallel.\cite{Snyder_NatMater2008} (b) To get the basic idea of $ZT$, one may use the simplified single-element model. [See Section~(\ref{sec:ZT})] }\label{fig:ZT}
    \end{figure}

            Besides $G$, $S$, and $K_e$, the figure of merit, $ZT$, is another important quantity for thermoelectric materials. $ZT$ is a dimensionless number determining the maximum thermoelectric energy conversion efficiency. As shown in Figure~\ref{fig:ZT}(a), a thermoelectric device usually consists of $\Pi$-shape junctions, which are made of $n$-type and $p$-type thermoelectric materials in serial. To get a basic idea of $ZT$, we consider a simplified model as drawn in Figure~\ref{fig:ZT}(b), where a thermoelectric element with different temperatures at two ends, $T_{h,c}$, serves as a thermoelectric generator and supplies power directly to a resistor with resistance $R$ in the closed circuit. The power supply comes from the Seebeck effect, which leads to the open-circuit voltage of $\Delta V=S\Delta T$ ($\Delta T=T_h-T_c$). Thus, heat is converted to work through the Seebeck effect. How much is the energy conversion efficiency in this circuit?

            We can naturally define the conversion efficiency as the ratio of work over heat
            \begin{align}\label{ZT:eta}
            \eta  = \frac{W}{Q}.
            \end{align}
            The work done in the circuit is the Joule heat
            \begin{align}\label{ZT:W}
            W = {I^2}R.
            \end{align}
            At the same time, the heat extracted from the hot end consists of the Fourier heat and Peltier heat from the hot end; the heat added up to the hot end is half of the Joule heat generated inside the thermoelectric element. As a result, we get
            \begin{align}\label{ZT:Q}
            Q = K \Delta T + STI -{I^2}r/2,
            \end{align}
            where $r=1/G$ is the resistance of the thermoelectric element, $K$ contains thermal conductance of electrons and phonons, and
            \begin{align}\label{ZT:I}
            I = \frac{{S\Delta T}}{{R + r}}.
            \end{align}
            Substituting Equations~(\ref{ZT:W})-(\ref{ZT:I}) into Equation~(\ref{ZT:eta}), and defining
            \begin{align}
            Z = \frac{{{S^2}}}{{K r}},\\
            m=R/r,
            \end{align}
            we obtain the formula for heat-electricity conversion efficiency:
            \begin{align}
            \eta  = \frac{{\Delta T}}{{{T_h}}}\frac{{\frac{m}{{m + 1}}}}{{1 + \frac{{m + 1}}{{Z{T_h}}} - \frac{1}{{2\left( {m + 1} \right)}}\frac{{\Delta T}}{{{T_h}}}}}.
            \end{align}
            This formula shows shows that the energy conversion efficiency varies with $m$. The maximum efficiency can be obtained by     letting $\partial{\eta}/\partial{m}=0$, which results in
            \begin{align}
            m = \sqrt {1 + ZT},
            \end{align}
            where
            \begin{align}
            T=(T_c+T_h)/2,
            \end{align}
            and the maximum energy conversion efficiency as
            \begin{align}\label{eq:eta}
            \eta_{\textrm{max}}  = \frac{{\Delta T}}{T_h}\frac{{\sqrt {1 + ZT}  - 1}}{{{T_c}/{T_h} + \sqrt {1 + ZT} }}.
            \end{align}
            The prefactor $\frac{{\Delta T}}{T_h}$ is the value for the Carnot cycle and also the upper limit of energy conversion efficiency. Therefore, the maximum energy conversion is determined by the dimensionless quantity $ZT$.\cite{Ioffe_1957}

            If only the thermal conduction of electrons and phonons is taken into account, $K=K_e+K_{ph}$, and
            \begin{align}
            ZT = \frac{{G{S^2}T}}{{{K _e} + {K _{ph}}}}.
            \end{align}
            It is worth noting that in bulk materials, the electrical conductance $G$ and thermal conductance $K$ can be replaced by electrical conductivity and thermal conductivity, respectively.\cite{Snyder2012} Most importantly, higher $ZT$ guarantees larger maximum heat-work conversion efficiency.

            To evaluate the conversion efficiency, one may do a quick calculation. For example, when $T_h=400$~K, $T_c=300$~K, and $ZT=4$, we get from Equation (\ref{eq:eta}) a maximum heat-energy conversion efficiency of $\eta\approx 10\%$, which is close to geothermal heat-power conversion efficiency.\cite{Vining_NatMater2009} Researchers have searched for high-$ZT$ materials for decades. However, for bulk 3D materials, the $ZT$ is generally less than or around 1. For example, alloys of Bi$_2$Te$_3$ and Sb$_2$Te$_3$, which are widely-used as thermoelectric materials, have peak $ZT$ values typically in the range of 0.8 to 1.\cite{Snyder_NatMater2008} As proposed by Hicks and Dresselhaus in 1993, $ZT$ can be significantly improved in low-dimensional materials.\cite{Hick_1993}

        \subsubsection{Spin-dependent linear response theory\label{ZcsT}}
            Besides the charge current, the temperature gradient may induce spin current in magnetic materials or structures, such as ferromagnetic Ni$_{81}$Fe$_{19}$ film attached with a Pt wire at one end. This phenomenon was discovered and named the Spin Seebeck effect in 2008.\cite{Uchida_nat2008} This effect was initially attributed to thermoelectric effects of individual spin channels of electrons with unreasonably-long spin coherent length and was later interpreted as magnon-driven spin Seebeck effects.\cite{XiaoJiang_PRB2010} In 2012, G. E. W. Bauer suggested naming the former effects caused by spin-dependent transport of electrons as ``spin-dependent'' Seebeck effects.\cite{Bauer_NatMater_2012,ZhaohuiThesis}

            In the following, we shall introduce the basic ideas of spin-dependent linear response theory. For simplicity, we ignore the spin dependence of temperature, which is actually effectively quenched by interspin and electron-phonon scattering.\cite{Bauer_NatMater_2012} When the spin degree of freedom is involved, the linear response equation [Equation~(\ref{eq:IIQ})] can be extended to\cite{cxb_prb2013_ac,Bauer_NatMater_2012}
            \begin{align}\label{eq:spinLinear}
            \left( {\begin{array}{*{20}{l}}
            I\\
            {I_s}\\
            {{I_Q}}
            \end{array}} \right) = \left( {\begin{array}{*{20}{c}}
            {{G_c}}&{{G_s}}&{{\Pi _c}}\\
            {{G_s}}&{{G_c}}&{{\Pi _s}}\\
            {e\Pi {_c}}&{e{\Pi _s}}&{K}
            \end{array}} \right)
            \left( {\begin{array}{*{20}{l}}
            \Delta V\\
            {\Delta {V_s}}/2\\
            {\Delta T}
            \end{array}} \right),
            \end{align}
            where the charge/spin electrical conductance $G_{c/s}$ and the Seebeck coefficient $S_{c/s}$ are
            \begin{align}
            G_{c/s}&=G_{\uparrow} \pm G_{\downarrow}, \\
            S_c&=(S_{\uparrow}+S_{\downarrow})/2,\\
            S_s&=S_{\uparrow}-S_{\downarrow}.
            \end{align}
            Herein, the spin-resolved/charge/spin Peltier coefficient $\Pi _{\sigma/c/s }$, thermal conductance $K$, charge voltage $\Delta V$, and spin voltage $\Delta V_s$ are defined as
            \begin{align}
            {\Pi _\sigma } & = {G_\sigma }{S_\sigma },\\
            \Pi {  _c} &= \Pi {  _ \uparrow } + \Pi {  _ \downarrow },\\
            \Pi {  _s} &= \Pi {  _ \uparrow } - \Pi {  _ \downarrow },\\
            K&={{K _e} + T\left( {S_ \uparrow ^2{G_ \uparrow } + S_ \downarrow ^2{G_ \downarrow }} \right)},\\
            e\Delta V &= \Delta \mu=\left( {{\mu _{L \uparrow }} + {\mu _{L \downarrow }} - {\mu _{R \uparrow }} - {\mu _{R \downarrow }}} \right)/2,\\
            e\Delta {V_s} &= \Delta \mu_s=\left( {{\mu _{L \uparrow }} - {\mu _{L \downarrow }} - {\mu _{R \uparrow }} + {\mu _{R \downarrow }}} \right) .
            \end{align}
            And the spin-resolved charge current is obtained as
            \begin{align}\label{eq:Iupdown}
             I_{\uparrow,\downarrow} &= \frac{{e}}  {h}\int { \mathcal{T}_{\uparrow,\downarrow} \left( E \right)\left( { f_{FD;L}  - f_{FD;R}} \right){\textrm d} E }.
            \end{align}
            When electronic spin currents are thermally induced, heat is converted to work as well. Thus, energy conversion efficiency can be defined similarly. In order to investigate the energy conversion efficiency, several kinds of formulas with slight differences in the charge Seebeck coefficient $S_c$, spin voltage $\mu_s$, and charge/spin figure of merit $Z_{c/s}T$ have been proposed.\cite{Dubi_prb2009,Swirkowicz_PRB_2009,Rejec_prb_2012,cxb_prb2013_ac}

            \textbf{Charge figure of merit.} Due to the complexity introduced by one extra driving force originating from spin voltage $V_s$, it is not straightforward to obtain a simple expression for energy conversion efficiency. Generally, there are two methods to roughly estimate the energy efficiency.

            The first method ignores the spin voltage:
            \begin{align}
            V_s=0,
            \end{align}
            such that
            \begin{align}
            I =  {G_c}\Delta V + {\Pi _c}\Delta T,
            \end{align}
            according to Equation~(\ref{eq:spinLinear}). Thus, under the situation $V_s=0$ and $I=0$,
            \begin{align}\label{eq:Sc1}
            { S_c} \equiv {\left( { - \frac{{\Delta V}}{{\Delta T}}} \right)_{I = 0,{V_s} = 0}} = \frac{{{G_ \uparrow }{S_ \uparrow } + {G_ \downarrow }{S_ \downarrow }}}{{{G_ \uparrow } + {G_ \downarrow }}}.
            \end{align}
            This formula is similar to the definition of the Seebeck coefficient defined in semiconductors with bipolar effects.\cite{Goldsmid_book}
            The electrical conductance is obtained by setting the temperature gradient equal to zero:
            \begin{align}
            G_c=I/\Delta V=G_\uparrow+G_\downarrow.
            \end{align}

            The second method ignores the spin current:
            \begin{align}
            I_s=0,
            \end{align}
            such that $S_c$ is defined under the situation of an open-circuit condition, where both charge and spin currents are zero.\cite{cxb_prb2013_ac,cxb_PRB_2014} Using Equation~(\ref{eq:spinLinear}), we know that
            \begin{align}\left\{ \begin{array}{cll}
            I &= {G_c}\Delta V + \frac{1}{2}{G_s}\Delta {V_s} + {\Pi _c}\Delta T = 0,\\
            {I_s} &= {G_s}\Delta V + \frac{1}{2}{G_c}\Delta {V_s} + {\Pi _s}\Delta T = 0,
            \end{array} \right.\end{align}
            which leads to
            \begin{align}
            {S_c} \equiv {\left( { - \frac{{\Delta V}}{{\Delta T}}} \right)_{I = 0,{I_s} = 0}} =(S_\uparrow+S_\downarrow)/2.
            \end{align}
            In both cases, the charge figure of merit is approximately defined as
             \begin{align}
            Z_cT=G_cS_c^2T/K.
            \end{align}

            \textbf{Spin figure of merit.} To assess the energy conversion efficiency of heat to spin current, an analogy is made to define the spin figure of merit:\cite{Dubi_prb2009,Swirkowicz_PRB_2009,cxb_prb2013_ac}
            \begin{align}
            Z_sT=G_sS_s^2T/K,
            \end{align}
            where
            \begin{align}
            G_s&=G_\uparrow-G_\downarrow,\\
            S_s&=S_\uparrow-S_\downarrow.
            \end{align}
            Similarly, $Z_sT$ is utilized in the search for high-efficiency heat-spin conversion materials.

        \subsubsection{Valley-dependent linear response theory}
            ``Valley'' is used to label equal-energy, but inequivalent, states in $k$-space. For example, graphene has two energetically degenerate, but inequivalent, valleys near the Fermi energy at the the corners of the Brillouin Zone,\cite{XiaoDi_PRL2007} and bulk silicon has six degenerate and equivalent minima of the conduction band.\cite{Takashina_PRL_2006}
            As another degree of freedom, valley is promising in its future application for the storage and manipulation of information.
            A crucial requirement for Valleytronics is a long valley life-time. The lifetime of valley excitons in monolayer transition-metal dichalcogenides typically ranges from a few to hundreds of picoseconds.\cite{XuXiaodong_NatPhys2014,MaCong_NanoLett2014} Lifetimes exceeding 3 nanoseconds at 5 K were observed in electron-doped MoS$_2$ and WS$_2$ monolayers.\cite{YangLuyi_NatPhys2015} Recently, even-longer lifetimes of microseconds were realized in WSe$_2$/MoS$_2$ heterostructures.\cite{WangFeng_SciAdv2017}

            When electronic transport is valley-dependent, the valley-resolved charge current is
            \begin{equation}
                       I_{\eta} = \frac{{e}}  {h}\int { \mathcal{T}_{\eta} \left( E \right)\left(  f_{FD;L}  - f_{FD;R} \right){\textrm d} E },
            \end{equation}
            where $\eta$ labels different valleys.
            Within linear response approximation, the above equation can be written as a linear function of valley-dependent electrical conductance and the Seebeck coefficient ($G_\eta$ and $S_\eta$) as
            \begin{equation}\label{eq:valleyI}
            I_\eta=G_\eta \Delta V + G_\eta S_\eta \Delta T.
            \end{equation}
            The definitions of $G_\eta$ and $S_\eta$ in a spin-degenerate system are \cite{cxb_PRB2015valley}
            \begin{align}
            {G_\eta }& = \frac{{2{e^2}}}{h}\int_{}^{} {{\mathcal{T}_\eta }\left( E \right){{\left. {\left( { - {\partial _E} f_{FD}} \right)} \right|}_{T,\mu }}dE} ,\\
            {S_\eta } &=  - \frac{1}{{|e|T}}\frac{{\int_{}^{} {{\mathcal{T}_\eta }\left( E \right)\left( {E - \mu } \right){{\left. {\left( {{\partial _E}f_{FD}} \right)} \right|}_{{T},{\mu }}}dE} }}{{\int_{}^{} {{\mathcal{T} _\eta }\left( E \right){{\left. {\left( {{\partial _E}f_{FD}} \right)} \right|}_{{T},{\mu }}}dE} }}\ , \label{eq:S}
            \end{align}
            respectively. Here, ${\mathcal{T} _\eta }(E)$ is the electronic transmission function of $\eta$-valley. Unlike Equation (\ref{eq:spinLinear}), there are only two driving forces considered in this case. In spin-split systems, spin index is included, resembling Equation (\ref{eq:Iupdown}).

    \subsection{Thermoelectric devices}
    To investigate thermoelectric properties of nano devices, an analysis of both transmission functions of electrons and phonons are needed. With the knowledge of the transmission functions of electrons and the Seebeck coefficient, the electrical conductance and the electronic thermal conductance can be obtained using Equations~(\ref{G})-(\ref{k}) for normal thermoelectric properties. Also, with the transmission function of phonons, phonon thermal conductance can be computed using Equation~(\ref{eq:thermConductance}). For pristine systems, one can directly use the ballistic transmission function, which is equal to the number of carrier channels at the given energy. For quasi-ballistic systems with structural defects, transmission functions of electrons and phonons can be numerically calculated using the electron and phonon NEGF methods, respectively.

    Nanostructures and low-dimensional materials provide new opportunities for achieving higher-$ZT$-value materials and devices.\cite{Dresselhaus_AM2007,Dirmyer2009,HanGuang_small2014,YangRonggui_2016,ZhangQingjie_small2016} As mentioned before, the $ZT$ of 1D silicon nanowires with rough surfaces has a 100-fold increase of the $ZT$ due to the reduction of thermal conductivity.\cite{YangPeidong_2008}
    Zou \emph{et~al.} made further investigation into the thermoelectric properties of thin GaAs nanowires. Although bulk GaAs is poor in thermoelectricity, the room-temperature $ZT$ of an ideal GaAs nanowire with a diameter of 1.1 nm was predicted to reach up to 1.34, which was also a 100-fold improvement over the bulk GaAs.\cite{Zou_nanoscale2015} More enhancement could be achieved by introducing surface doping for roughness to further reduce thermal conductance.\cite{Zou_nanoscale2015}

    Basing their research on the $\pi$-orbital tight-binding model and the Brenner potential, Gunst $et~al.$ investigated finite-size graphene anti-dot structures and found a maximum room-temperature $ZT$ of 0.25. They showed that the thermoelectric performance of graphene antidot structures is highly sensitive to antidot edges.\cite{Gunst_PRB_2011} By using the NEGF methods for both the electrons and phonons as well as a fourth-nearest-neighbor force constant model\cite{SaitoBook,Zimmermann_PRB2008} for constructing the dynamical matrix, Chen $et~al$ studied thermoelectric properties of graphene nanoribbons, junctions, and superlattices.\cite{ChenY_JPCM2010} It was shown that the maximum $ZT$ was largely controlled by the narrowest part of a junction due to the restriction of phonon thermal conductance by the narrowest part,\cite{Xuyong_prb_2010} and that increasing the number of interfaces enhanced the peak $ZT$ value. Also, a maximum value 0.63 of $ZT$ was expected for chevron-type graphene nanoribbons,\cite{ChenY_JPCM2010}
    and high $ZT$ values were also expected in anti-ferromagnetic silicene nanoribbons.\cite{Zberecki_PRB2013}
    In addition, a Boron nitride quantum dot structure may have a 2 to 3 times larger $ZT$ than the same carbon structures,\cite{LongMengqiu_ResInPhys2017} and hybrid graphene/boron nitride nanoribbons exhibit a $ZT$ of 1.5$\sim$3 times larger.\cite{YangKaike_PRB2012}
    It is also promising to enhance $ZT$ in molecular junctions.\cite{Natalya_JPCM2017} For example, connecting a spin-nondegenerate orbital level to two leads, Ren $et~al.$ showed that by increasing electron-phonon coupling and Coulomb repulsion, $ZT$ could be enhanced.\cite{RenJie_PRB2012}

\begin{figure}
      \centering
      \includegraphics[width=0.6 \textwidth]{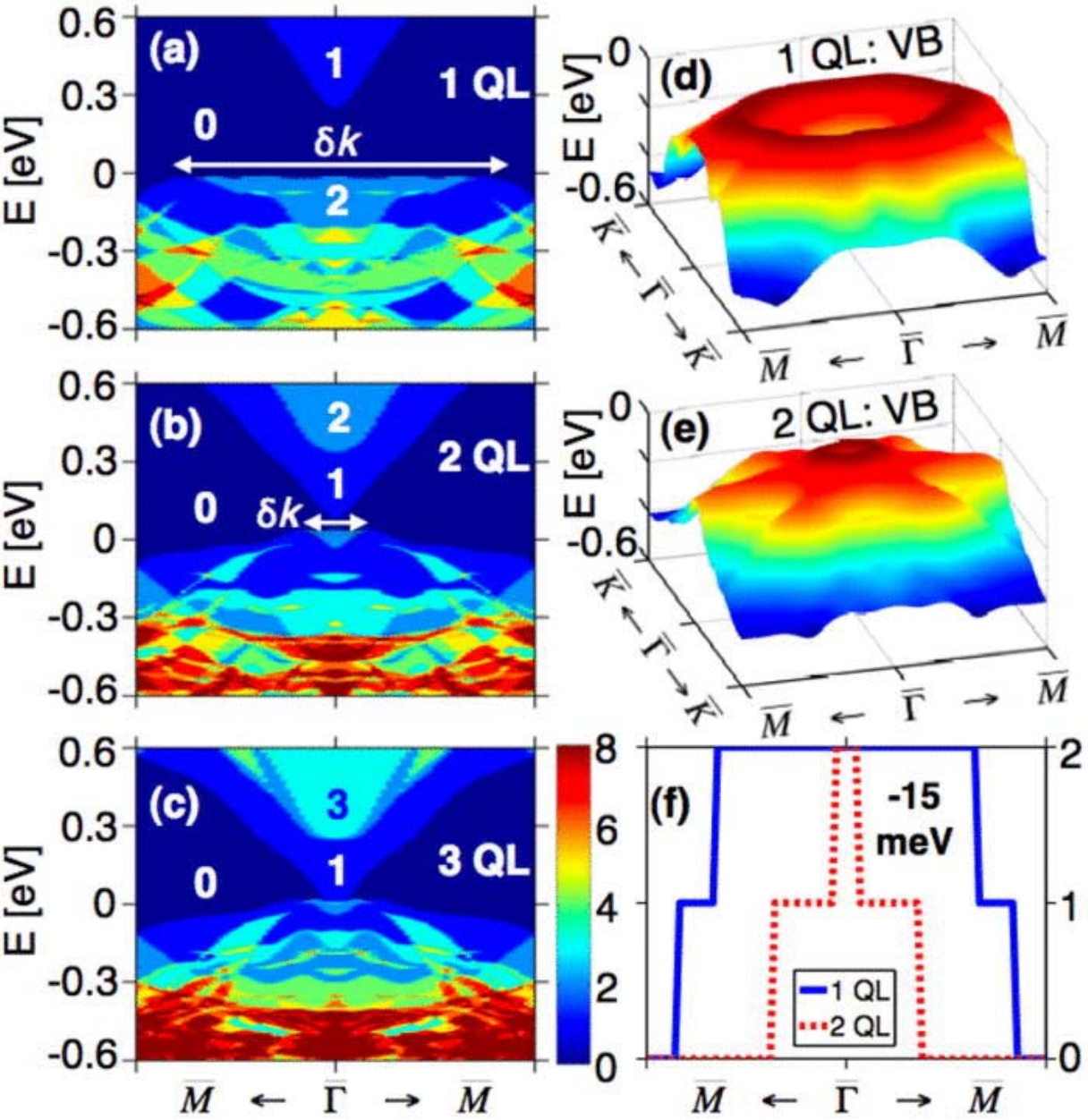}\\
      \caption{(a)-(c) Transmission evolution of Bi$_2$Se$_3$ as the number of layers changes from 1 QL to 3 QL. (d)(e) Surface plots for valence bands for 1 QL and 2QL Bi$_2$Se$_3$. (f) Transmission at an energy of -15 meV below the top of the valence band. Reproduced with permission.\cite{Jesse_APL2013} Copyright 2013, AIP Publishing.}\label{fig8}
    \end{figure}

    Besides the above quasi-1D structures, 2D materials, especially topological materials, are of great interest.\cite{Xuyong_PRL2014,SiChen_topo2014,SiChen_topo2016,Huaqing_review2017} However, for 2D devices with sizes exceeding the phonon mean free path, electron-phonon scattering should be taken into consideration.
    Zahid \emph{et~al.} studied the thermoelectric performance of Bi$_2$Te$_3$ quintuple (QL) thin films with phonon-phonon scattering introduced to the effective electronic transmission function $\bar {\mathcal{T}} \left( \omega  \right)$ as
    \begin{align}
    \bar {\mathcal{T}} \left( \omega  \right) =\mathcal{T} \left( \omega  \right)\frac{{{\lambda}}}{L},
    \end{align}
    where $\mathcal{T}\left( \omega  \right)$ is the density of modes (transmission of the ideal structure per unit area), $\lambda$ is the electron MFP, and $L$ is the device length.\cite{Zahid_apl_2010} When $\lambda$ is dominated by electron-phonon scattering, the energy dependence of $\lambda$ can be ignored. In Zahid $et~al.$'s work, the value of $\lambda$ was obtained by fitting the Seebeck coefficient of bulk Bi$_2$Te$_3$ as a function of chemical potential to the experimental value. By using an experimental value of thermal conductivity as 1.5 W K$^{-1}$m$^{-1}$, they found a maximum value of $ZT=7.15$ for a 2D crystalline Bi$_2$Te$_3$ quintuple thin film, which had a 10-fold increase compared to the bulk value. This increase was attributed to the change of valence bands due to quantum confinement in thin films. Later, Jesse $et~al.$ investigated the thermoelectric performance of few-layer Bi$_2$Te$_3$ with thickness ranging from 1 to 6 quintuple layers (Figure~\ref{fig8}).\cite{Jesse_APL2013} A constant $\lambda$ was assumed for both valence and conduction bands such that the value $\lambda$ did not influence the results of $S$. They found that the thinnest film exhibits the best MFP-scaled power factor ($GS^2/\lambda$) and attributed this enhancement to its particular shape of valence bands, which was closely related to the interaction of topological surface states.

    Although Bi$_2$Se$_3$ has similar structures and properties to Bi$_2$Te$_3$, experimental measurements of few-layer Bi$_2$Se$_3$ performed by Guo $et~al.$ showed a remarkable decrease of the power factor ($GS^2$) when the thickness is reduced from 30 to 5 quintuple layers, differing the aforementioned theoretical predictions for Bi$_2$Te$_3$.\cite{GuoMinghua_NJP2016} Because of the existence of topological surface states, the Seebeck coefficient is now constituted by two parts:
    \begin{align}
    S=(\sigma_s S_s +\sigma_b S_b)/(\sigma_s  +\sigma_b ),
    \end{align}
    where $S_{s/b}(\sigma_{s/b})$ is the Seebeck coefficient (electrical conductivity) for surface/bulk states. Considering that gapless surface states emerge when the number of quintuple layers exceeds about six,\cite{Kyungwha_PRL2010} the difference may be caused by the variation of surface states, the electronic mean free path as the thickness varies, and the Fermi energy in the experiments. The connection, however, between topological insulators and thermoelectric materials is still unclear and being investigated because of its serious research and development potential.

    \subsection{Spin caloritronic devices}
    Spintronics concerns the active manipulation of spin degrees of freedom in solid-state systems.\cite{Zutic_RMP2004}
    Spin-polarized current and pure spin current are key factors for spintronics since they can change the magnetic orientation of a ferromagnet without the application of a magnetic field.\cite{cxb_PRB2017_STT} Thermal manipulation of spintronics is called ``spin caloritronics" and was introduced in 2012 to represent the burgeoning research field that focuses on the interaction of spin with heat currents.\cite{Bauer_NatMater_2012}

    Spin caloritronic phenomena can be categorized into three classes:\cite{Bauer_NatMater_2012} (i) effects caused by the spin-dependent conduction of electrons; (ii) effects due to the collective dynamics of the magnetic order parameter; (ii) effects related to the relativistic spin-orbit coupling, such as thermal generalization of Hall effects.\cite{ZhangLifa_PRL2010}
    Since the discovery of ``spin Seebeck effect'' by Uchida $et~al.$ in 2008,\cite{Uchida_nat2008} 
    there has been a burst of investigation on spin caloritronic effects.\cite{Uchida_NatMat_2010insulator,XiaoJiang_PRB2010}

\begin{figure}
       \centering
      \includegraphics[width=0.6\textwidth]{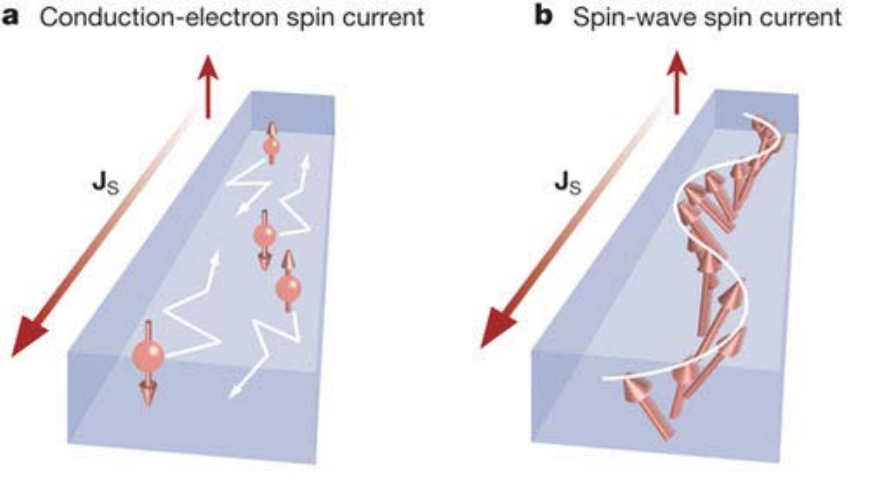}\\
      \caption{Spin currents can be carried by electrons or magnons: spin angular momentum $\bar J_S$ is carried by (a) conducting electrons, and (b) collective magnetic-moment precession. Reprinted by permission.\cite{Kajiwara_Nat2010} Copyright 2001, Macmillan Publishers Ltd.
      }\label{fig:spinCurrent}
    \end{figure}

    Spin caloritronics introduces many intriguing factors to consider, such as the giant magneto-Seebeck effect, anomalously large thermal spin transfer torque,\cite{Hatami_PRL_2007,zhaohui} 100\%-spin-polarized current, and pure spin current.\cite{ZengMinggang_NanoLett2011,Rejec_prb_2012,cxb_PRB_2014}  
    As shown in Figure~\ref{fig:spinCurrent}, spin currents can be carried by conducting electrons or by magnons. Pure spin currents can be easily realized with the aid of spin-dependent thermoelectric effect,\cite{ZengMinggang_NanoLett2011,Rejec_prb_2012,cxb_PRB_2014} where a pure spin current can be generated by simply varying the chemical potential, or via the anomalous Nernst effect in monolayer Group-VI dichalcogenides when the Fermi level is lying in the valence band.\cite{Jauho_PRL2015}

    The spin-dependent thermoelectric effects, as briefly discussed in Section~\ref{ZcsT}, are natural consequences of spin-dependent electronic transport, such as that observed in ferromagnetic materials. Experimental detections of the spin-dependent Seebeck effect\cite{Slachter_NatPhy_2010} and Peltier effect\cite{Flipse_2012} were reported in spin valve structures in 2010 and 2012, respectively. A giant spin-dependent Seebeck effect was shown in magnetic tunnel junctions, where the charge Seebeck coefficient exceeds 1 mV/K.\cite{linWW_NatComm_2012}
    Furthermore, the spin-dependent Seebeck effect is also discussed in research on silicene or graphene-based magnetic junctions,\cite{NiuZhiping_PLA2014,OuyangTao_JAP_2015} molecular junctions,\cite{LiJianwei_PRB2016} Rashba quantum dot system,\cite{JiangFeng_PLA_2014} \textit{etc.} 

    Although $S_s$ is only $-2$ nV/K in permalloy Ni$_{81}$Fe$_{19}$,\cite{Uchida_nat2008} theoretical investigation shows that $S_s$ reaches about 50 $\mu$V/K for a free electronic system with Rashba spin-orbit interaction and external magnetic field,\cite{LuHaifeng_APL2010}
    growing as large as 3.4 mV/K in spin semiconductors.\cite{cxb_PRB_2014}

    The spin-dependent Seebeck effect also drives thermal spin transfer torque.\cite{Choi_NatPhys2015} It is predicted that thermally generated spin transfer torques can be anomalously large.\cite{Hatami_PRL_2007} Some indications manifest in experiments,\cite{yu,zhaohui} yet solid evidence remains to be found.

    \subsection{Valley caloritronic devices}
    Valley polarization,\cite{Takashina_PRL_2006} polarized valley current, and pure valley current are being pursued in the studies of valleytronics.
    To acquire valley polarization, one can apply strain,\cite{Gunawan_PRL_2006} electric field,\cite{Ezawa_PRL2012} optical selective excitation,\cite{YaoWang_PRB_2008,FengJi_NatCommun_2012} magnetic field,\cite{Takashina_PRL_2006} and temperature gradient.\cite{NiuZP_APL_2014,cxb_PRB2015valley} To obtain a valley-polarized current, one can utilize
    valley filtering by carefully-designed twisting junctions in SiC nanoribbons;\cite{Zheng_2D2017}
    use circularly polarized light in monolayer transition-metal dichalcogenides;\cite{lei,YuYujin_Nanotech2016} adopt the negative refraction of $n$-$p$-$n$ junctions;\cite{Garcia_PRL2008} or tune the gate,\cite{XuFM_NJP2016} the incident energy,\cite{YangMou_PRB2016} and scattering potential barriers.\cite{WangJing_JPCM2016}

    Extended line defects,\cite{Gunlycke_PRL_2011,CTWhite_NanoLett_2012} wedge-shaped graphene nanoribbons,\cite{Beenakker_NatPhys_2007,cxb_PRB2015valley} and graphene nanobubbles\cite{Jauho_PRL2016} also act as valley filters. Finally, to realize a pure valley current, which means that there is no accompanying net charge current, one can apply adiabatic cyclic strain field\cite{Jiang_PRL_2013} and quantum pumping.\cite{WangJing_APE2014}

    In light of valleytronics and analogous to spin caloritronics,\cite{Bauer_NatMater_2012,cxb_PRB_2014} the concept of ``Valley caloritronics'' has also been proposed.\cite{cxb_PRB2015valley} Generally, the basic idea of valley caloritronics is to thermally initiate valley current---either a valley-polarized or a pure valley current.

\begin{figure}
          \centering
      \includegraphics[width=0.6\textwidth]{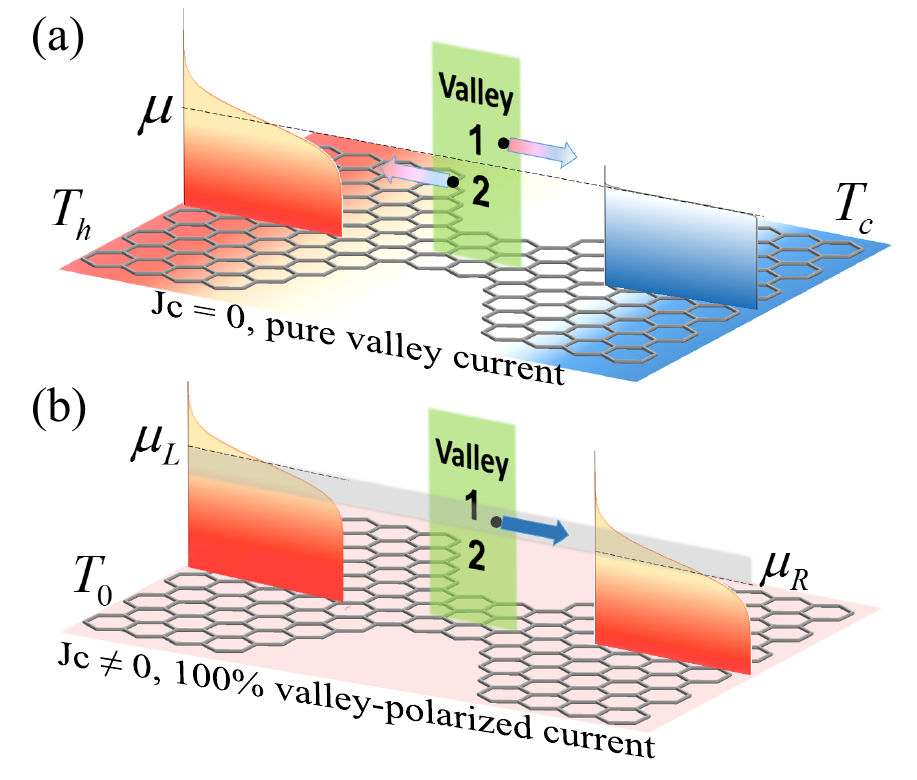}\\
      \caption{Basic idea of the generation of a pure valley current in valley caloritronic devices. (a) Under a temperature bias, electrons of different valleys move in opposite directions if their Seebeck voltages $\Delta V^\eta_T=G_\eta S_\eta\Delta T $ have different signs. (b) Under a voltage bias, electrons within the energy window [$\mu_R$,$\mu_L$] of both valleys move in the same direction. Reproduced with permission.\cite{cxb_PRB2015valley} Copyright 2015, American Physical Society.
     }\label{fig:valley}
    \end{figure}

    As shown in Figure~\ref{fig:valley}, the advantage of thermal initiation is that a pure valley current can be generated in a two-probe system, which is impossible using a dc bias voltage. The key point lies in the sign-variable Seebeck coefficients, which determine the valley-dependent thermoelectric current flow according to Equation (\ref{eq:valleyI}) as\cite{cxb_PRB2015valley}
    \begin{align}
    {I_\eta } = {G_\eta }{S_\eta }\Delta T.
    \end{align}
    Because the valley-dependent Seebeck coefficient $S_\eta$ can be negative or positive, and because the valley-dependent conductance $G_\eta$ is always positive, we know that the direction of the valley-dependent current is determined by $S_\eta$. As a consequence, the total charge current $I=\sum_\eta I_\eta$ can be zero, implying a pure valley current. Compared to electrical methods,\cite{Jiang_PRL_2013,WangJing_APE2014} thermal generation of a pure valley currents is straight-forward and easier to realize.

    Besides in zigzag graphene nanoribbons,\cite{cxb_PRB2015valley,YuZhizhou_Carbon2016} thermal generation of a pure valley current is also proposed in ferromagnetic silicene junctions\cite{NiuZP_APL_2014,ZhaiXC_PRB2016} and in group-IV monolayers\cite{ZhaiXC_NJP2017} under the application of an external magnetic field.

    By analogy with thermal manipulation of electronic valleys, manipulation of phonon valleys is also possible. Yet, to the best of our knowledge, this field has not yet been investigated.

\section{Summary and outlook}
    Accompanying the development of nano devices and experimental techniques, anomalous thermal transport properties are discovered in nanoscale systems. Similar to the NEGF method for electronic transport, the phonon NEGF method proves to be effective for exploring the quantum transport properties of nanoscale systems. In this review, we first gave a detailed formalism of the phonon NEGF method and presented investigations on quantum thermal transport in nanoscale systems, which covered thermal transport in nanojunctions, anisotropy of thermal conduction, and the impacts of point defects and strain. We also briefly discussed the phonon topology and the extended phonon NEGF method in topological phononic systems.
    Second, we introduced the linear response theory and its application in spin-dependent and valley-dependent systems. Then, we further illustrated this line of research by showing recent studies using the NEGF methods on thermoelectric and spin/valley caloritronic devices, which are manipulated by applying a temperature gradient.

    For future directions of research on nanoscale thermal engineering, there are two main objectives. The first is the manipulation of thermal conductance/conductivity in realistic materials, which includes:
    (1) controlling the magnitude of thermal conductance, especially by finding materials or structures that have extremely-high thermal conductance for the best heat dissipation, or extremely-low thermal conductance for optimized thermoelectric performance;
    (2) controlling other quasi-particles such as magnons, skyrmions, Cooper pairs, topological surface states, \textit{etc.} by varying the temperature.
    Discoveries of new materials, especially topological ones, open opportunities for both thermoelectrics and spin/valley caloritronics.\cite{XuYong_CPB} The second is the development of NEGF methods, such as the study of transient quantum thermal transport \cite{WangJiansheng_PRB2010} and the combination of NEGF methods with density functional theory to investigate large-size systems.\cite{Vincent2016} We hope that this review will inspire more discoveries on quantum effects of thermal transport and more developments on thermal engineering of low-dimensional devices.

\textbf{Acknowledgments}
We would like to thank John Allaster and Danni Cai from McGill University for their assistance with the language of this manuscript. This work was supported by the Ministry of Science and Technology of China (Grant No. 2016YFA0301001), and the National Natural Science Foundation of China (Grant Nos. 11704257, 11674188, and 11334006).

\textbf{Conflict of Interest}
The authors declare no conflict of interest.


\begin{thebibliography}{100}

\bibitem{LiBaowen_RMP_2012}
N.~Li, J.~Ren, L.~Wang, G.~Zhang, P.~H\"anggi, and B.~Li, \emph{Rev. Mod.
  Phys.}  \textbf{2012}, \emph{84}, 1045.

\bibitem{ZhangGang_book}
G.~Zhang, \emph{Nanoscale Energy Transport and Harvesting: A Computational
  Study}, Pan Stanford Publishing, Singapore \textbf{2013}.

\bibitem{Cahill_JAP_2003}
D.~G. Cahill, W.~K. Ford, K.~E. Goodson, G.~D. Mahan, A.~Majumdar, H.~J. Maris,
  R.~Merlin, and S.~R. Phillpot, \emph{J. Appl. Phys.}  \textbf{2003},
  \emph{93}, 793.

\bibitem{Pop_IEEE_2006}
E.~Pop, S.~Sinha, and K.~Goodson, \emph{Proc. IEEE.}  \textbf{2006}, \emph{94},
  1587.

\bibitem{Graham_small2005}
A.~P. Graham, G.~S. Duesberg, R.~V. Seidel, M.~Liebau, E.~Unger, W.~Pamler,
  F.~Kreupl, and W.~Hoenlein, \emph{Small}  \textbf{2005}, \emph{1}, 382.

\bibitem{LiBaowen_AdvMat2013}
J.~Wang, L.~Zhu, J.~Chen, B.~Li, and J.~T.~L. Thong, \emph{Adv. Mater.}
  \textbf{2013}, \emph{25}, 6884.

\bibitem{Balandin_NanoLett2008}
A.~A. Balandin, S.~Ghosh, W.~Bao, I.~Calizo, D.~Teweldebrhan, F.~Miao, and
  C.~N. Lau, \emph{Nano Lett.}  \textbf{2008}, \emph{8}, 902.

\bibitem{Xuyong_small_2014}
Y.~Xu, Z.~Li, and W.~Duan, \emph{Small}  \textbf{2014}, \emph{10}, 2182.

\bibitem{YaPeidong_2008}
A.~Hochbaum, R.~Chen, R.~Delgado, W.~Liang, E.~Garnett, M.~Najarian,
  A.~Majumdar, and P.~Yang, \emph{Nature}  \textbf{2008}, \emph{451}, 163.

\bibitem{Majumda}
D.~Li, Y.~Wu, P.~Kim, L.~Shi, P.~Yang, and A.~Majumdar, \emph{Appl. Phys.
  Lett.}  \textbf{2003}, \emph{83}, 2934.

\bibitem{Zettl_PRL_2008}
C.~Chang, D.~Okawa, H.~Garcia, A.~Majumdar, and A.~Zettl, \emph{Phys. Rev.
  Lett.}  \textbf{2008}, \emph{101}, 75903.

\bibitem{LiBaowen_NatComm_2014}
X.~Xu, L.~F.~C. Pereira, Y.~Wang, J.~Wu, K.~Zhang, X.~Zhao, S.~Bae,
  C.~Tinh~Bui, R.~Xie, J.~T.~L. Thong, B.~H. Hong, K.~P. Loh, D.~Donadio,
  B.~Li, and B.~\"Oyilmaz, \emph{Nat. Commun.}  \textbf{2014}, \emph{5}, 3689.

\bibitem{Pendry_1983}
J.~Pendry, \emph{J. Phys. A: Math. Gen.}  \textbf{1983}, \emph{16}, 2161.

\bibitem{Maynard_PRB_1985}
R.~Maynard and E.~Akkermans, \emph{Phys. Rev. B}  \textbf{1985}, \emph{32},
  5440.

\bibitem{Angelescu_1998}
D.~Angelescu, M.~Cross, and M.~Roukes, \emph{Superlattice. Microstruct.}
  \textbf{1998}, \emph{23}, 673.

\bibitem{Rego_PRL_1998}
L.~G.~C. Rego and G.~Kirczenow, \emph{Phys. Rev. Lett.}  \textbf{1998},
  \emph{81}, 232.

\bibitem{Blencowe_PRB_1999}
M.~P. Blencowe, \emph{Phys. Rev. B}  \textbf{1999}, \emph{59}, 4992.

\bibitem{Schwab_nat_2000}
K.~Schwab, E.~Henriksen, J.~Worlock, and M.~Roukes, \emph{Nature}
  \textbf{2000}, \emph{404}, 974.

\bibitem{HoneJ_Sci_2000}
J.~Hone, B.~Batlogg, Z.~Benes, A.~Johnson, and J.~Fischer, \emph{Science}
  \textbf{2000}, \emph{289}, 1730.

\bibitem{Adu_NanoLett2005}
K.~Adu, H.~Gutierrez, U.~Kim, G.~Sumanasekera, and P.~Eklund, \emph{Nano Lett.}
   \textbf{2005}, \emph{5}, 409.

\bibitem{Breton_Nat_2011}
J.~Le~Breton, S.~Sharma, H.~Saito, S.~Yuasa, and R.~Jansen, \emph{Nature}
  \textbf{2011}, \emph{475}, 82.

\bibitem{cxb_PRB_2014}
X.~Chen, Y.~Liu, B.-L. Gu, W.~Duan, and F.~Liu, \emph{Phys. Rev. B}
  \textbf{2014}, \emph{90}, 121403(R).

\bibitem{XiaoJiang_PRB2010}
J.~Xiao, G.~E.~W. Bauer, K.~Uchida, E.~Saitoh, and S.~Maekawa, \emph{Phys. Rev.
  B}  \textbf{2010}, \emph{81}, 214418.

\bibitem{Cheng_PRL2016}
R.~Cheng, S.~Okamoto, and D.~Xiao, \emph{Phys. Rev. Lett.}  \textbf{2016},
  \emph{117}, 217202.

\bibitem{TangGaomin_arxiv2017}
G.~Tang, X.~Chen, J.~Ren, and J.~Wang, \emph{Phys. Rev. B} 
  \textbf{2018}, \emph{97}, 081407(R).

\bibitem{cxb_PRB2015valley}
X.~Chen, L.~Zhang, and H.~Guo, \emph{Phys. Rev. B}  \textbf{2015}, \emph{92},
  155427.

\bibitem{YangRonggui_PRB2004}
R.~Yang and G.~Chen, \emph{Phys. Rev. B}  \textbf{2004}, \emph{69}, 195316.

\bibitem{LiWu_PRB2015}
W.~Li, N.~Mingo, L.~Lindsay, D.~A. Broido, D.~A. Stewart, and N.~A. Katcho,
  \emph{Phys. Rev. B}  \textbf{2012}, \emph{85}, 195436.

\bibitem{LiWu_CPC2016}
W.~Li, J.~Carrete, N.~A. Katcho, and N.~Mingo, \emph{Comput. Phys. Commun.}
  \textbf{2014}, \emph{185}, 1747.

\bibitem{ZhangGang_JCP_2005}
G.~Zhang and B.~Li, \emph{J. Chem. Phys.}  \textbf{2005}, \emph{123}, 114714.

\bibitem{chenYP_NanoLett_2009}
J.~Hu, X.~Ruan, and Y.~Chen, \emph{Nano Lett.}  \textbf{2009}, \emph{9}, 2730.

\bibitem{GongXG_APL2009}
Z.~Guo, D.~Zhang, and X.-G. Gong, \emph{Appl. Phys. Lett.}  \textbf{2009},
  \emph{95}, 163103.

\bibitem{Keblinski_APL2010}
W.~J. Evans, L.~Hu, and P.~Keblinski, \emph{Appl. Phys. Lett.}  \textbf{2010},
  \emph{96}, 203112.

\bibitem{Pop_PRB2011}
Z.-Y. Ong and E.~Pop, \emph{Phys. Rev. B}  \textbf{2011}, \emph{84}, 075471.

\bibitem{ZhangGang_Nanoscale2011}
G.~Zhang and H.~Zhang, \emph{Nanoscale}  \textbf{2011}, \emph{3}, 4604.

\bibitem{Cagin_ACSNano2011}
J.~Haskins, A.~Kinaci, C.~Sevik, H.~Sevin\c{c}li, G.~Cuniberti, and
  T.~\c{C}a\v{g}in, \emph{ACS Nano}  \textbf{2011}, \emph{5}, 3779.

\bibitem{XuBaoxing_Carbon2016}
Y.~Gao, W.~Yang, and B.~Xu, \emph{Carbon}  \textbf{2016}, \emph{96}, 513.

\bibitem{ChenChaoguang_Carbon2017}
S.-K. Chien and Y.-T. Yang, \emph{Carbon}  \textbf{2012}, \emph{50}, 421.

\bibitem{WangJS_EPJ_2008}
J.~Wang, J.~Wang, and J.~L¨¹, \emph{Eur. Phys. J. B}  \textbf{2008}, \emph{62},
  381.

\bibitem{Xuyong_APL2009}
Y.~Xu, X.~Chen, B.-L. Gu, and W.~Duan, \emph{Appl. Phys. Lett.}  \textbf{2009},
  \emph{95}, 233116.

\bibitem{Xuyong_prb_2010}
Y.~Xu, X.~Chen, J.-S. Wang, B.-L. Gu, and W.~Duan, \emph{Phys. Rev. B}
  \textbf{2010}, \emph{81}, 195425.

\bibitem{ZhuHongqin_NJP2012}
H.~Zhu, Y.~Xu, B.~Gu, and W.~Duan, \emph{New J. Phys.}  \textbf{2012},
  \emph{14}, 013053.

\bibitem{cxb_PUCNT}
X.~Chen, Y.~Xu, X.~Zou, B.-L. Gu, and W.~Duan, \emph{Phys. Rev. B}
  \textbf{2013}, \emph{87}, 155438.

\bibitem{Huaqing_PRB_2013}
H.~Huang, Y.~Xu, X.~Zou, J.~Wu, and W.~Duan, \emph{Phys. Rev. B}
  \textbf{2013}, \emph{87}, 205415.

\bibitem{ChenKeqiu_SciRep2015}
X.-F. Peng and K.-Q. Chen, \emph{Sci. Rep.}  \textbf{2015}, \emph{5}, 16215.

\bibitem{GuoZengyuan2015}
Y.~Dong and Z.-Y. Guo, \emph{J. Nanosci. Nanotechnol.}  \textbf{2015},
  \emph{15}, 3229.

\bibitem{Ziman_book}
J.~M. Ziman, \emph{Electrons and phonons: the theory of transport phenomena in
  solids}, Clarendon Press Oxford, UK \textbf{2001}.

\bibitem{wang_prb_2006}
J.-S. Wang, J.~Wang, and N.~Zeng, \emph{Phys. Rev. B}  \textbf{2006},
  \emph{74}, 033408.

\bibitem{yamamoto_PRL2006}
T.~Yamamoto and K.~Watanabe, \emph{Phys. Rev. Lett.}  \textbf{2006}, \emph{96},
  255503.

\bibitem{Jauho_Book}
H.~Haug and A.-P. Jauho, \emph{Quantum kinetics in transport and optics of
  semiconductors}, Springer-Verlag, Berlin \textbf{1996}.

\bibitem{WangJiansheng_PRE2007}
J.~Wang, N.~Zeng, J.~Wang, and C.~Gan, \emph{Phys. Rev. E}  \textbf{2007},
  \emph{75}, 61128.

\bibitem{WangJS_FoP2014}
J.-S. Wang, B.~K. Agarwalla, H.~Li, and J.~Thingna, \emph{Front. Phys.}
  \textbf{2014}, \emph{9}, 673.

\bibitem{Mingo_PRB_2003GF}
N.~Mingo and L.~Yang, \emph{Phys. Rev. B}  \textbf{2003}, \emph{68}, 245406.

\bibitem{note2}
The full Keldysh equation is $G^ < =
  (1+G^r\Sigma^r)G_0^<(1+\Sigma^aG^a)+G^r\Sigma^<G^a$ (The subscript ``CC'' for
  $G^\gamma$ is omitted here.)\cite{Jauho_Book} The second term is contributed
  by bound states and generally has no contribution to conductance.

\bibitem{Meir_PRL1992}
Y.~Meir and N.~S. Wingreen, \emph{Phys. Rev. Lett.}  \textbf{1992}, \emph{68},
  2512.

\bibitem{CaroliFormula_JPCS1971}
C.~Caroli, R.~Combescot, P.~Nozieres, and D.~Saint-James, \emph{J. Phys. C}
  \textbf{1971}, \emph{4}, 916.

\bibitem{Jauho_PRB_1994}
A.-P. Jauho, N.~S. Wingreen, and Y.~Meir, \emph{Phys. Rev. B}  \textbf{1994},
  \emph{50}, 5528.

\bibitem{mingo_prb_2006}
N.~Mingo, \emph{Phys. Rev. B}  \textbf{2006}, \emph{74}, 125402.

\bibitem{Morooka_PRB2008}
M.~Morooka, T.~Yamamoto, and K.~Watanabe, \emph{Phys. Rev. B}  \textbf{2008},
  \emph{77}, 033412.

\bibitem{LiBaowen_small2012}
C.~T. Bui, R.~Xie, M.~Zheng, Q.~Zhang, C.~H. Sow, B.~Li, and J.~T.~L. Thong,
  \emph{Small}  \textbf{2012}, \emph{8}, 738.

\bibitem{ZhangGang_EPJB2012}
S.~Liu, X.~F. Xu, R.~G. Xie, G.~Zhang, and B.~W. Li, \emph{Eur. Phys. J. B}
  \textbf{2012}, \emph{85}, 337.

\bibitem{yamamoto_prl_2004}
T.~Yamamoto, S.~Watanabe, and K.~Watanabe, \emph{Phys. Rev. Lett.}
  \textbf{2004}, \emph{92}, 075502.

\bibitem{Cui_science2017}
L.~Cui, W.~Jeong, S.~Hur, M.~Matt, J.~C. Kl\"ockner, F.~Pauly, P.~Nielaba,
  J.~C. Cuevas, E.~Meyhofer, and P.~Reddy, \emph{Science}  \textbf{2017},
  \emph{355}, 1192.

\bibitem{Mingo_PRB_2003dispersion}
N.~Mingo, \emph{Phys. Rev. B}  \textbf{2003}, \emph{68}, 113308.

\bibitem{Lundstrom_JAP_2010}
C.~Jeong, R.~Kim, M.~Luisier, S.~Datta, and M.~Lundstrom, \emph{J. Appl. Phys.}
   \textbf{2010}, \emph{107}, 023707.

\bibitem{Buttiker_PRB1985}
M.~B\"uttiker, Y.~Imry, R.~Landauer, and S.~Pinhas, \emph{Phys. Rev. B}
  \textbf{1985}, \emph{31}, 6207.

\bibitem{Dattabook}
S.~Datta, \emph{Electronic transport in mesoscopic systems}, Cambridge
  University Press, UK \textbf{1997}.

\bibitem{OuyangTao_PRB2017}
T.~Ouyang, Y.~Chen, Y.~Xie, X.~L. Wei, K.~Yang, P.~Yang, and J.~Zhong,
  \emph{Phys. Rev. B}  \textbf{2010}, \emph{82}, 245403.

\bibitem{WangJian_APL2012}
J.~Wang and J.-S. Wang, \emph{Appl. Phys. Lett.}  \textbf{2006}, \emph{88},
  111909.

\bibitem{CallawayBook}
J.~Callaway, \emph{Quantum theory of the solid state}, Academic Press, 2nd
  edition \textbf{1991}.

\bibitem{Wang_PRB2009}
J.~Wang and H.~Guo, \emph{Phys. Rev. B}  \textbf{2009}, \emph{79}, 045119.

\bibitem{xuyongPhD}
Y.~Xu, \emph{Quantum Thermal Transport in Nanoscale Systems}, Ph.D. thesis,
  Tsinghua University, Beijing \textbf{2010}.

\bibitem{Mingo_PRB2008}
N.~Mingo, D.~A. Stewart, D.~A. Broido, and D.~Srivastava, \emph{Phys. Rev. B}
  \textbf{2008}, \emph{77}, 1098.

\bibitem{Dhar_PRL_2008}
A.~Dhar and J.~L. Lebowitz, \emph{Phys. Rev. Lett.}  \textbf{2008}, \emph{100},
  134301.

\bibitem{XuYong_PRB2008}
Y.~Xu, J.-S. Wang, W.~Duan, B.-L. Gu, and B.~Li, \emph{Phys. Rev. B}
  \textbf{2008}, \emph{78}, 224303.

\bibitem{TanZW_NanoLett_2011}
Z.~Tan, J.~Wang, and C.~Gan, \emph{Nano Lett.}  \textbf{2011}, \emph{11}, 214.

\bibitem{Zou_nanoscale2015}
X.~Zou, X.~Chen, H.~Huang, Y.~Xu, and W.~Duan, \emph{Nanoscale}  \textbf{2015},
  \emph{7}, 8776.

\bibitem{SaitoBook}
R.~Saito, G.~Dresselhaus, and M.~Dresselhaus, \emph{Physical properties of
  carbon nanotubes}, World Scientific \textbf{1998}.

\bibitem{Sancho1985}
M.~Sancho, J.~Sancho, J.~Sancho, and J.~Rubio, \emph{J. Phys. F: Met. Phys.}
  \textbf{1985}, \emph{15}, 851.

\bibitem{LiBaowen2017}
S.~Hu, M.~An, N.~Yang, and B.~Li, \emph{Small}  \textbf{2017}, \emph{13},
  1602726.

\bibitem{XuBS_PDAP2016}
Z.~Yang, Y.-L. Ji, G.~Lan, L.-C. Xu, H.~Wang, X.~Liu, and B.~Xu, \emph{J. Phys.
  D: Appl. Phys.}  \textbf{2016}, \emph{49}, 145102.

\bibitem{HuJiuning_APL2010}
J.~Hu, S.~Schiffli, A.~Vallabhaneni, X.~Ruan, and Y.~P. Chen, \emph{Appl. Phys.
  Lett.}  \textbf{2010}, \emph{97}, 133107.

\bibitem{Balandin_NatMat_2011}
A.~Balandin, \emph{Nat. Mater.}  \textbf{2011}, \emph{10}, 569.

\bibitem{Balandin_JPCM2012}
D.~L. Nika and A.~A. Balandin, \emph{J. Phys.: Condens. Matter}  \textbf{2012},
  \emph{24}, 233203.

\bibitem{PoP_MrsBull_2012}
E.~Pop, V.~Varshney, and A.~K. Roy, \emph{MRS. Bull.}  \textbf{2012},
  \emph{37}, 1273.

\bibitem{RuanXiulin_NanoMicro2014}
Y.~Wang, A.~K. Vallabhaneni, B.~Qiu, and X.~Ruan, \emph{Nanosc. Microsc.
  Therm.}  \textbf{2014}, \emph{18}, 155.

\bibitem{QiminYan_nanolett_2007}
Q.~Yan, B.~Huang, J.~Yu, F.~Zheng, J.~Zang, J.~Wu, B.-L. Gu, F.~Liu, and
  W.~Duan, \emph{Nano Lett.}  \textbf{2007}, \emph{7}, 1469.

\bibitem{Cui_JCP2017}
L.~Cui, R.~Miao, C.~Jiang, E.~Meyhofer, and P.~Reddy, \emph{J. Chem. Phys.}
  \textbf{2017}, \emph{146}, 092201.

\bibitem{Klocker_PRB_2016}
J.~C. Kl\"ockner, M.~B\"urkle, J.~C. Cuevas, and F.~Pauly, \emph{Phys. Rev. B}
  \textbf{2016}, \emph{94}, 205425.

\bibitem{Klockner_PRB2017}
J.~C. Kl\"ockner, R.~Siebler, J.~C. Cuevas, and F.~Pauly, \emph{Phys. Rev. B}
  \textbf{2017}, \emph{95}, 245404.

\bibitem{Jinhao_InSe2017}
H.~Jin, J.~Li, L.~Wan, Y.~Dai, Y.~Wei, and H.~Guo, \emph{2D Mater.}
  \textbf{2017}, \emph{4}, 025116.

\bibitem{YangJK_small2011}
J.~Yang, Y.~Yang, S.~W. Waltermire, T.~Gutu, A.~A. Zinn, T.~T. Xu, Y.~Chen, and
  D.~Li, \emph{Small}  \textbf{2011}, \emph{7}, 2334.

\bibitem{Chen_APL2009}
Z.~Chen, W.~Jang, W.~Bao, C.~N. Lau, and C.~Dames, \emph{Appl. Phys. Lett.}
  \textbf{2009}, \emph{95}, 183.

\bibitem{Freitag_NanoLett2009}
M.~Freitag, M.~Steiner, Y.~Martin, V.~Perebeinos, Z.~Chen, J.~C. Tsang, and
  P.~Avouris, \emph{Nano Lett.}  \textbf{2009}, \emph{9}, 1883.

\bibitem{Jauregui_ECStran2010}
L.~A. Jauregui, Y.~Yue, A.~N. Sidorov, J.~Hu, Q.~Yu, G.~Lopez, R.~Jalilian,
  D.~K. Benjamin, D.~A. Delkd, and W.~Wu, \emph{ECS Trans.}  \textbf{2010},
  \emph{28}, 73.

\bibitem{Mingo_PRB2007}
W.~Zhang, N.~Mingo, and T.~Fisher, \emph{Phys. Rev. B}  \textbf{2007},
  \emph{76}, 195429.

\bibitem{Fisher_APL2016}
K.~Miao, S.~Sadasivam, J.~Charles, G.~Klimeck, T.~Fisher, and T.~Kubis,
  \emph{Appl. Phys. Lett.}  \textbf{2016}, \emph{108}, 113107.

\bibitem{Fisher_PRB2017}
S.~Sadasivam, N.~Ye, J.~P. Feser, J.~Charles, K.~Miao, T.~Kubis, and T.~S.
  Fisher, \emph{Phys. Rev. B}  \textbf{2017}, \emph{95}, 085310.

\bibitem{ZouXL_AccChem2015}
X.~Zou and B.~I. Yakobson, \emph{Accounts. Chem. Res.}  \textbf{2015},
  \emph{48}, 73.

\bibitem{NanoLett2013_XZou}
X.~Zou, Y.~Liu, and B.~I. Yakobson, \emph{Nano Lett.}  \textbf{2013},
  \emph{13}, 253.

\bibitem{ZouXL_small2015}
X.~Zou and B.~I. Yakobson, \emph{Small}  \textbf{2015}, \emph{11}, 4503.

\bibitem{ChenKeqiu_Carbon2013}
S.-H. Tan, L.-M. Tang, Z.-X. Xie, C.-N. Pan, and K.-Q. Chen, \emph{Carbon}
  \textbf{2013}, \emph{65}, 181.

\bibitem{LanJinhua_JPDAP2014}
J.~Lan, Y.~Cai, G.~Zhang, J.-S. Wang, and Y.-W. Zhang, \emph{J. Phys. D: Appl.
  Phys.}  \textbf{2014}, \emph{47}, 265303.

\bibitem{Huaqing_JPCM2015}
H.~Huang, \emph{J. Phys.: Condens. Matter}  \textbf{2015}, \emph{27}, 305402.

\bibitem{Markussen_nanoLett2008}
T.~Markussen, A.-P. Jauho, and M.~Brandbyge, \emph{Nano Lett.}  \textbf{2008},
  \emph{8}, 3771.

\bibitem{OuyangTao_Nanotech_2010}
T.~Ouyang, Y.~Chen, Y.~Xie, K.~Yang, Z.~Bao, and J.~Zhong,
  \emph{Nanotechnology}  \textbf{2010}, \emph{21}, 245701.

\bibitem{ChenYuanping_Nanotech2016}
Z.~Zhang, Y.~Xie, Q.~Peng, and Y.~Chen, \emph{Nanotechnology}  \textbf{2016},
  \emph{27}, 445703.

\bibitem{OuyangTao_PRB_2012}
T.~Ouyang, Y.~Chen, L.-M. Liu, Y.~Xie, X.~Wei, and J.~Zhong, \emph{Phys. Rev.
  B}  \textbf{2012}, \emph{85}, 235436.

\bibitem{OuyangTao_Nanotech2014}
O.~Tao and H.~Ming, \emph{Nanotechnology}  \textbf{2014}, \emph{25}, 245401.

\bibitem{LiDengfeng_APL_2014}
D.~Li, Y.~Xu, X.~Chen, B.~Li, and W.~Duan, \emph{Appl. Phys. Lett.}
  \textbf{2014}, \emph{104}, 143108.

\bibitem{LiDengfeng2013}
D.~Li, B.~Li, M.~Luo, C.~Feng, T.~Ouyang, and F.~Gao, \emph{Appl. Phys. Lett.}
  \textbf{2013}, \emph{103}, 071908.

\bibitem{ZhangYongwei_JPCC2014}
Z.-Y. Ong, Y.~Cai, G.~Zhang, and Y.-W. Zhang, \emph{J. Phys. Chem. C}
  \textbf{2014}, \emph{118}, 25272.

\bibitem{WuJunqiao_NatComm2015}
S.~Lee, F.~Yang, J.~Suh, S.~Yang, Y.~Lee, G.~Li, H.~S. Choe, A.~Suslu, Y.~Chen,
  and C.~Ko, \emph{Nat. Commun.}  \textbf{2015}, \emph{6}, 8573.

\bibitem{ShiLi2016}
B.~Smith, B.~Vermeersch, J.~Carrete, E.~Ou, J.~Kim, N.~Mingo, D.~Akinwande, and
  L.~Shi, \emph{Adv. Mater.}  \textbf{2016}, \emph{2}, 1603756.

\bibitem{ZhangGang_PRB2016}
H.~Zhou, Y.~Cai, G.~Zhang, and Y.-W. Zhang, \emph{Phys. Rev. B}  \textbf{2016},
  \emph{94}, 045423.

\bibitem{ZhuBangfen_PRB_2005}
Y.-Q. Cheng, S.-Y. Zhou, and B.-F. Zhu, \emph{Phys. Rev. B}  \textbf{2005},
  \emph{72}, 035410.

\bibitem{Mingo_PRL_2008isotope}
I.~Savic, N.~Mingo, and D.~A. Stewart, \emph{Phys. Rev. Lett.}  \textbf{2008},
  \emph{101}, 165502.

\bibitem{Jauho_PRL_2009}
T.~Markussen, A.-P. Jauho, and M.~Brandbyge, \emph{Phys. Rev. Lett.}
  \textbf{2009}, \emph{103}, 055502.

\bibitem{ChenKeqiu_PRB2010}
X.-F. Peng, K.-Q. Chen, Q.~Wan, B.~Zou, and W.~Duan, \emph{Phys. Rev. B}
  \textbf{2010}, \emph{81}, 195317.

\bibitem{PengXF_Carbon2014}
X.-F. Peng and K.-Q. Chen, \emph{Carbon}  \textbf{2014}, \emph{77}, 360.

\bibitem{ZhangGang_MechMater2015}
G.~Zhang and Y.-W. Zhang, \emph{Mech. Mater.}  \textbf{2015}, \emph{91}, 382.

\bibitem{Lindsay_PRB2014}
L.~Lindsay, W.~Li, J.~Carrete, N.~Mingo, D.~A. Broido, and T.~L. Reinecke,
  \emph{Phys. Rev. B}  \textbf{2014}, \emph{89}, 155426.

\bibitem{GongXingao_APL2009}
Z.~Guo, D.~Zhang, and X.-G. Gong, \emph{Appl. Phys. Lett.}  \textbf{2009},
  \emph{95}, 163103.

\bibitem{SiChen_Nano2016}
C.~Si, Z.~Sun, and F.~Liu, \emph{Nanoscale}  \textbf{2016}, \emph{8}, 3207.

\bibitem{Menglei2015}
M.~Li, J.~Li, L.-Q. Chen, B.-L. Gu, and W.~Duan, \emph{Phys. Rev. B}
  \textbf{2015}, \emph{92}, 115435.

\bibitem{zhoumei}
M.~Zhou, X.~Chen, M.~Li, and A.~Du, \emph{J. Mater. Chem. C}  \textbf{2017},
  \emph{5}, 1247.

\bibitem{zhouMei2015}
M.~Zhou, W.~Duan, Y.~Chen, and A.~Du, \emph{Nanoscale}  \textbf{2015},
  \emph{7}, 15168.

\bibitem{Jinhao_strain2017}
H.~Jin, J.~Li, Y.~Dai, and Y.~Wei, \emph{Phys. Chem. Chem. Phys.}
  \textbf{2017}, \emph{19}, 4855.

\bibitem{JinGuojun_EPL2011}
X.~Zhai and G.~Jin, \emph{EPL}  \textbf{2011}, \emph{96}, 16002.

\bibitem{PengXF_Carbon2016}
X.-F. Peng, K.-Q. Chen, X.-J. Wang, and S.-H. Tan, \emph{Carbon}
  \textbf{2016}, \emph{100}, 36.

\bibitem{LuJunqiang_PRL2003}
J.-Q. Lu, J.~Wu, W.~Duan, F.~Liu, B.-F. Zhu, and B.-L. Gu, \emph{Phys. Rev.
  Lett.}  \textbf{2003}, \emph{90}, 156601.

\bibitem{Liu2017b}
Y.~Liu, Y.~Xu, and W.~Duan, \emph{Natl. Sci. Rev.}
  \textbf{2017}, nwx086.

\bibitem{Xiao2010}
D.~Xiao, M.-C. Chang, and Q.~Niu, \emph{Rev. Mod. Phys.}  \textbf{2010},
  \emph{82}, 1959.

\bibitem{Liu2017a}
Y.~Liu, Y.~Xu, S.-C. Zhang, and W.~Duan, \emph{Phys. Rev. B}  \textbf{2017},
  \emph{96}, 064106.

\bibitem{ZhangLifa_PRL2010}
L.~Zhang, J.~Ren, J.-S. Wang, and B.~Li, \emph{Phys. Rev. Lett.}
  \textbf{2010}, \emph{105}, 225901.

\bibitem{susstrunk2016}
R.~S{\"u}sstrunk and S.~D. Huber, \emph{Proc. Natl. Acad. Sci. USA}
  \textbf{2016}, \emph{113}, E4767.

\bibitem{Liu2017c}
Y.~Liu, C.-S. Lian, Y.~Li, Y.~Xu, and W.~Duan, \emph{Phys. Rev. Lett.}  \textbf{2017}, \emph{119}, 255901.

\bibitem{Nash2015}
L.~M. Nash, D.~Kleckner, A.~Read, V.~Vitelli, A.~M. Turner, and W.~T. Irvine,
  \emph{Proc. Natl. Acad. Sci. U.S.A}  \textbf{2015}, \emph{112}, 14495.

\bibitem{Susstrunk2015}
R.~S{\"u}sstrunk and S.~D. Huber, \emph{Science}  \textbf{2015}, \emph{349},
  47.

\bibitem{he2016}
C.~He, X.~Ni, H.~Ge, X.-C. Sun, Y.-B. Chen, M.-H. Lu, X.-P. Liu, and Y.-F.
  Chen, \emph{Nat. Phys.}  \textbf{2016}, \emph{12}, 1124.

\bibitem{peng2016}
Y.-G. Peng, C.-Z. Qin, D.-G. Zhao, Y.-X. Shen, X.-Y. Xu, M.~Bao, H.~Jia, and
  X.-F. Zhu, \emph{Nat. Commun.}  \textbf{2016}, \emph{7}, 13368.

\bibitem{Chen2014}
B.~G.-g. Chen, N.~Upadhyaya, and V.~Vitelli, \emph{Proc. Natl. Acad. Sci.
  U.S.A.}  \textbf{2014}, \emph{111}, 13004.

\bibitem{Paulose2015a}
J.~Paulose, A.~S. Meeussen, and V.~Vitelli, \emph{Proc. Natl. Acad. Sci.
  U.S.A.}  \textbf{2015}, \emph{112}, 7639.

\bibitem{Paulose2015b}
J.~Paulose, B.~G.-g. Chen, and V.~Vitelli, \emph{Nat. Phys.}  \textbf{2015},
  \emph{11}, 153.

\bibitem{Chen2016}
B.~G.-g. Chen, B.~Liu, A.~A. Evans, J.~Paulose, I.~Cohen, V.~Vitelli, and
  C.~Santangelo, \emph{Phys. Rev. Lett.}  \textbf{2016}, \emph{116}, 135501.

\bibitem{Meeussen2016}
A.~S. Meeussen, J.~Paulose, and V.~Vitelli, \emph{Phys. Rev. X}  \textbf{2016},
  \emph{6}, 041029.

\bibitem{MahanBook}
G.~D. Mahan, \emph{Many-particle physics}, Kluwer Academic / Plenum Publishers,
  New York, 3rd edition \textbf{2000}.

\bibitem{Ioffe_1957}
A.~Ioffe, \emph{Semiconductor thermoelements, and Thermoelectric cooling},
  Infosearch Ltd., London \textbf{1957}.

\bibitem{Snyder2012}
Y.~Pei, H.~Wang, and G.~J. Snyder, \emph{Adv. Mater.}  \textbf{2012},
  \emph{24}, 6125.

\bibitem{Vining_NatMater2009}
C.~B. Vining, \emph{Nat. Mater.}  \textbf{2009}, \emph{8}, 83.

\bibitem{Snyder_NatMater2008}
G.~J. Snyder and E.~S. Toberer, \emph{Nat. Mater.}  \textbf{2008}, \emph{7},
  105.

\bibitem{Hick_1993}
L.~D. Hicks and M.~S. Dresselhaus, \emph{Phys. Rev. B}  \textbf{1993},
  \emph{47}, 12727.

\bibitem{Uchida_nat2008}
K.~Uchida, S.~Takahashi, K.~Harii, J.~Ieda, W.~Koshibae, K.~Ando, S.~Maekawa,
  and E.~Saitoh, \emph{Nature}  \textbf{2008}, \emph{455}, 778.

\bibitem{Bauer_NatMater_2012}
G.~E.~W. Bauer, E.~Saitoh, and B.~J. van Wees, \emph{Nat. Mater.}
  \textbf{2012}, \emph{11}, 391.

\bibitem{ZhaohuiThesis}
Z.~Zhang, \emph{Spin-dependent electrical and thermal transport in magnetic
  tunnel junctions}, Thesis, University of Manitoba \textbf{2016}.

\bibitem{cxb_prb2013_ac}
X.~Chen, D.~Liu, W.~Duan, and H.~Guo, \emph{Phys. Rev. B}  \textbf{2013},
  \emph{87}, 085427.

\bibitem{Dubi_prb2009}
Y.~Dubi and M.~Di~Ventra, \emph{Phys. Rev. B}  \textbf{2009}, \emph{79},
  081302.

\bibitem{Swirkowicz_PRB_2009}
R.~Swirkowicz, M.~Wierzbicki, and J.~Barnas, \emph{Phys. Rev. B}
  \textbf{2009}, \emph{80}, 195409.

\bibitem{Rejec_prb_2012}
T.~Rejec, R.~\v{Z}itko, J.~Mravlje, and A.~Ram\v{s}ak, \emph{Phys. Rev. B}
  \textbf{2012}, \emph{85}, 085117.

\bibitem{Goldsmid_book}
H.~J. Goldsmid, \emph{Introduction to thermoelectricity}, volume 121, Springer
  \textbf{2010}.

\bibitem{XiaoDi_PRL2007}
D.~Xiao, W.~Yao, and Q.~Niu, \emph{Phys. Rev. Lett.}  \textbf{2007}, \emph{99},
  236809.

\bibitem{Takashina_PRL_2006}
K.~Takashina, Y.~Ono, A.~Fujiwara, Y.~Takahashi, and Y.~Hirayama, \emph{Phys.
  Rev. Lett.}  \textbf{2006}, \emph{96}, 236801.

\bibitem{XuXiaodong_NatPhys2014}
X.~Xu, W.~Yao, D.~Xiao, and T.~F. Heinz, \emph{Nat. Phys.}  \textbf{2014},
  \emph{10}, 343.

\bibitem{MaCong_NanoLett2014}
C.~Mai, A.~Barrette, Y.~Yu, Y.~G. Semenov, K.~W. Kim, L.~Cao, and K.~Gundogdu,
  \emph{Nano Lett.}  \textbf{2014}, \emph{14}, 202.

\bibitem{YangLuyi_NatPhys2015}
L.~Yang, N.~A. Sinitsyn, W.~Chen, J.~Yuan, J.~Zhang, J.~Lou, and S.~Crooker,
  \emph{Nat. Phys.}  \textbf{2015}, \emph{11}, 830.

\bibitem{WangFeng_SciAdv2017}
J.~Kim, C.~Jin, B.~Chen, H.~Cai, T.~Zhao, P.~Lee, S.~Kahn, K.~Watanabe,
  T.~Taniguchi, S.~Tongay, M.~F. Crommie, and F.~Wang, \emph{Sci. Adv.}
  \textbf{2017}, \emph{3}, e1700518.

\bibitem{Dresselhaus_AM2007}
M.~S. Dresselhaus, G.~Chen, M.~Y. Tang, R.~G. Yang, H.~Lee, D.~Z. Wang, Z.~F.
  Ren, J.~P. Fleurial, and P.~Gogna, \emph{Adv. Mater.}  \textbf{2007},
  \emph{19}, 1043.

\bibitem{Dirmyer2009}
M.~R. Dirmyer, J.~Martin, G.~S. Nolas, A.~Sen, and J.~V. Badding, \emph{Small}
  \textbf{2009}, \emph{5}, 933.

\bibitem{HanGuang_small2014}
G.~Han, Z.-G. Chen, J.~Drennan, and J.~Zou, \emph{Small}  \textbf{2014},
  \emph{10}, 2747.

\bibitem{YangRonggui_2016}
X.~Gu and R.~Yang, \emph{Ann. Rev. Heat Transfer}  \textbf{2016}, \emph{19}, 1.

\bibitem{ZhangQingjie_small2016}
W.~Zhou, Q.~Fan, Q.~Zhang, K.~Li, L.~Cai, X.~Gu, F.~Yang, N.~Zhang, Z.~Xiao,
  H.~Chen, S.~Xiao, Y.~Wang, H.~Liu, W.~Zhou, and S.~Xie, \emph{Small}
  \textbf{2016}, \emph{12}, 3407.

\bibitem{YangPeidong_2008}
A.~Hochbaum, R.~Chen, R.~Delgado, W.~Liang, E.~Garnett, M.~Najarian,
  A.~Majumdar, and P.~Yang, \emph{Nature}  \textbf{2008}, \emph{451}, 163.

\bibitem{Gunst_PRB_2011}
T.~Gunst, T.~Markussen, A.-P. Jauho, and M.~Brandbyge, \emph{Phys. Rev. B}
  \textbf{2011}, \emph{84}, 155449.

\bibitem{Zimmermann_PRB2008}
J.~Zimmermann, P.~Pavone, and G.~Cuniberti, \emph{Phys. Rev. B}  \textbf{2008},
  \emph{78}, 045410.

\bibitem{ChenY_JPCM2010}
Y.~Chen, T.~Jayasekera, A.~Calzolari, K.~Kim, and M.~B. Nardelli, \emph{J.
  Phys.: Condens. Matter}  \textbf{2010}, \emph{22}, 372202.

\bibitem{Zberecki_PRB2013}
K.~Zberecki, M.~Wierzbicki, J.~Barna, and R.~Swirkowicz, \emph{Phys. Rev. B}
  \textbf{2013}, \emph{88}, 4673.

\bibitem{LongMengqiu_ResInPhys2017}
C.~Pan, M.~Long, and J.~He, \emph{Results Phys.}  \textbf{2017}, \emph{7},
  1487.

\bibitem{YangKaike_PRB2012}
K.~Yang, Y.~Chen, R.~D'Agosta, Y.~Xie, J.~Zhong, and A.~Rubio, \emph{Phys. Rev.
  B}  \textbf{2012}, \emph{86}, 045425.

\bibitem{Natalya_JPCM2017}
A.~Z. Natalya, \emph{J. Phys.: Condens. Matter}  \textbf{2016}, \emph{28},
  183002.

\bibitem{RenJie_PRB2012}
J.~Ren, J.-X. Zhu, J.~E. Gubernatis, C.~Wang, and B.~Li, \emph{Phys. Rev. B}
  \textbf{2012}, \emph{85}, 155443.

\bibitem{Xuyong_PRL2014}
Y.~Xu, Z.~Gan, and S.-C. Zhang, \emph{Phys. Rev. Lett.}  \textbf{2014},
  \emph{112}, 226801.

\bibitem{SiChen_topo2014}
C.~Si, J.~Liu, Y.~Xu, J.~Wu, B.-L. Gu, and W.~Duan, \emph{Phys. Rev. B}
  \textbf{2014}, \emph{89}, 115429.

\bibitem{SiChen_topo2016}
C.~Si, K.-H. Jin, J.~Zhou, Z.~Sun, and F.~Liu, \emph{Nano Lett.}
  \textbf{2016}, \emph{16}, 6584.

\bibitem{Huaqing_review2017}
H.~Huang, Y.~Xu, J.~Wang, and W.~Duan, \emph{WIREs Comput. Mol. Sci.}
  \textbf{2017}, \emph{7}, e1296.

\bibitem{Zahid_apl_2010}
F.~Zahid and R.~Lake, \emph{Appl. Phys. Lett.}  \textbf{2010}, \emph{97},
  212102.

\bibitem{Jesse_APL2013}
J.~Maassen and M.~Lundstrom, \emph{Appl. Phys. Lett.}  \textbf{2013},
  \emph{102}, 093103.

\bibitem{GuoMinghua_NJP2016}
M.~Guo, Z.~Wang, Y.~Xu, H.~Huang, Y.~Zang, C.~Liu, W.~Duan, Z.~Gan, S.-C.
  Zhang, K.~He, X.~Ma, Q.~Xue, and Y.~Wang, \emph{New J. Phys.}  \textbf{2016},
  \emph{18}, 015008.

\bibitem{Kyungwha_PRL2010}
K.~Park, J.~Heremans, V.~Scarola, and D.~Minic, \emph{Phys. Rev. Lett.}
  \textbf{2010}, \emph{105}, 186801.

\bibitem{Zutic_RMP2004}
I.~\v{Z}uti\'c, J.~Fabian, and S.~Das~Sarma, \emph{Rev. Mod. Phys.}
  \textbf{2004}, \emph{76}, 323.

\bibitem{cxb_PRB2017_STT}
X.~Chen, C.~Zhou, Z.~Zhang, J.~Chen, X.~Zheng, L.~Zhang, C.-M. Hu, and H.~Guo,
  \emph{Phys. Rev. B}  \textbf{2017}, \emph{95}, 115417.

\bibitem{Uchida_NatMat_2010insulator}
K.~Uchida, J.~Xiao, H.~Adachi, J.~Ohe, S.~Takahashi, J.~Ieda, T.~Ota,
  Y.~Kajiwara, H.~Umezawa, H.~Kawai, G.~E.~W. Bauer, S.~Maekawa, and E.~Saitoh,
  \emph{Nat. Mater.}  \textbf{2010}, \emph{9}, 894.

\bibitem{Hatami_PRL_2007}
M.~Hatami, G.~E.~W. Bauer, Q.~Zhang, and P.~J. Kelly, \emph{Phys. Rev. Lett.}
  \textbf{2007}, \emph{99}, 066603.

\bibitem{zhaohui}
Z.~Zhang, L.~Bai, X.~Chen, H.~Guo, X.~L. Fan, D.~S. Xue, D.~Houssameddine, and
  C.~M. Hu, \emph{Phys. Rev. B}  \textbf{2016}, \emph{94}, 064414.

\bibitem{ZengMinggang_NanoLett2011}
M.~Zeng, Y.~Feng, and G.~Liang, \emph{Nano Lett.}  \textbf{2011}, \emph{11},
  1369.

\bibitem{Jauho_PRL2015}
X.-Q. Yu, Z.-G. Zhu, G.~Su, and A.~P. Jauho, \emph{Phys. Rev. Lett.}
  \textbf{2015}, \emph{115}, 246601.

\bibitem{Slachter_NatPhy_2010}
A.~Slachter, F.~Bakker, J.~Adam, and B.~Van~Wees, \emph{Nat. Phys.}
  \textbf{2010}, \emph{6}, 879.

\bibitem{Flipse_2012}
J.~Flipse, F.~L. Bakker, A.~Slachter, F.~K. Dejene, and B.~J. van Wees,
  \emph{Nat. Nano}  \textbf{2012}, \emph{7}, 166.

\bibitem{linWW_NatComm_2012}
W.~Lin, M.~Hehn, L.~Chaput, B.~Negulescu, S.~Andrieu, F.~Montaigne, and
  S.~Mangin, \emph{Nat. Commun.}  \textbf{2012}, \emph{3}, 744.

\bibitem{NiuZhiping_PLA2014}
Z.~P. Niu, \emph{Phys. Lett. A}  \textbf{2014}, \emph{378}, 73.

\bibitem{OuyangTao_JAP_2015}
B.~Zhou, B.~Zhou, Y.~Zeng, G.~Zhou, and T.~Ouyang, \emph{J. Appl. Phys.}
  \textbf{2015}, \emph{117}, 104305.

\bibitem{LiJianwei_PRB2016}
J.~W. Li, B.~Wang, F.~M. Xu, Y.~D. Wei, and J.~Wang, \emph{Phys. Rev. B}
  \textbf{2016}, \emph{93}.

\bibitem{JiangFeng_PLA_2014}
F.~Jiang, H.~Xie, and Y.~Yan, \emph{Phys. Lett. A}  \textbf{2014}, \emph{378},
  1854.

\bibitem{LuHaifeng_APL2010}
H.-F. L\"u, L.-C. Zhu, X.-T. Zu, and H.-W. Zhang, \emph{Appl. Phys. Lett.}
  \textbf{2010}, \emph{96}, 123111.

\bibitem{Choi_NatPhys2015}
G.-M. Choi, C.-H. Moon, B.-C. Min, K.-J. Lee, and D.~G. Cahill, \emph{Nat.
  Phys.}  \textbf{2015}, \emph{11}, 576.

\bibitem{yu}
H.~Yu, S.~Granville, D.~Yu, and J.~P. Ansermet, \emph{Phys. Rev. Lett.}
  \textbf{2010}, \emph{104}, 146601.

\bibitem{Gunawan_PRL_2006}
O.~Gunawan, Y.~Shkolnikov, K.~Vakili, T.~Gokmen, E.~De~Poortere, and
  M.~Shayegan, \emph{Phys. Rev. Lett.}  \textbf{2006}, \emph{97}, 186404.

\bibitem{Ezawa_PRL2012}
M.~Ezawa, \emph{Phys. Rev. Lett.}  \textbf{2012}, \emph{109}, 055502.

\bibitem{YaoWang_PRB_2008}
W.~Yao, D.~Xiao, and Q.~Niu, \emph{Phys. Rev. B}  \textbf{2008}, \emph{77},
  235406.

\bibitem{FengJi_NatCommun_2012}
T.~Cao, G.~Wang, W.~Han, H.~Ye, C.~Zhu, J.~Shi, Q.~Niu, P.~Tan, E.~Wang,
  B.~Liu, and J.~Feng, \emph{Nat. Commun.}  \textbf{2012}, \emph{3}, 887.

\bibitem{NiuZP_APL_2014}
Z.~P. Niu and S.~Dong, \emph{Appl. Phys. Lett.}  \textbf{2014}, \emph{104},
  202401.

\bibitem{Zheng_2D2017}
X.~Zheng, X.~Chen, L.~Zhang, L.~Xiao, S.~Jia, Z.~Zeng, and H.~Guo, \emph{2D
  Mater.}  \textbf{2017}, \emph{4}, 025013.

\bibitem{lei}
L.~Zhang, K.~Gong, J.~Chen, L.~Liu, Y.~Zhu, D.~Xiao, and H.~Guo, \emph{Phys.
  Rev. B}  \textbf{2014}, \emph{90}, 195428.

\bibitem{YuYujin_Nanotech2016}
Y.~Yu, Y.~Zhou, L.~Wan, B.~Wang, F.~Xu, Y.~Wei, and J.~Wang,
  \emph{Nanotechnology}  \textbf{2016}, \emph{27}, 185202.

\bibitem{Garcia_PRL2008}
J.~L. Garcia-Pomar, A.~Cortijo, and M.~Nieto-Vesperinas, \emph{Phys. Rev.
  Lett.}  \textbf{2008}, \emph{100}, 236801.

\bibitem{XuFM_NJP2016}
F.~M. Xu, Z.~Z. Yu, Y.~F. Ren, B.~Wang, Y.~D. Wei, and Z.~H. Qiao, \emph{New J.
  Phys.}  \textbf{2016}, \emph{18}, 113011.

\bibitem{YangMou_PRB2016}
M.~Yang, Y.-K. Bai, W.-L. Zhang, and R.-Q. Wang, \emph{Phys. Rev. B}
  \textbf{2016}, \emph{94}, 075433.

\bibitem{WangJing_JPCM2016}
W.~Jing, L.~Mengqiu, Z.~Wen-Sheng, H.~Yue, W.~Gaofeng, and K.~S. Chan, \emph{J.
  Phys.: Condens. Matter}  \textbf{2016}, \emph{28}, 285302.

\bibitem{Gunlycke_PRL_2011}
D.~Gunlycke and C.~T. White, \emph{Phys. Rev. Lett.}  \textbf{2011},
  \emph{106}, 136806.

\bibitem{CTWhite_NanoLett_2012}
D.~Gunlycke, S.~Vasudevan, and C.~T. White, \emph{Nano Lett.}  \textbf{2012},
  \emph{13}, 259.

\bibitem{Beenakker_NatPhys_2007}
A.~Rycerz, J.~Tworzyd{\l}o, and C.~Beenakker, \emph{Nat. Phys.}  \textbf{2007},
  \emph{3}, 172.

\bibitem{Jauho_PRL2016}
M.~Settnes, S.~R. Power, M.~Brandbyge, and A.-P. Jauho, \emph{Phys. Rev. Lett.}
   \textbf{2016}, \emph{117}, 276801.

\bibitem{Jiang_PRL_2013}
Y.~Jiang, T.~Low, K.~Chang, M.~I. Katsnelson, and F.~Guinea, \emph{Phys. Rev.
  Lett.}  \textbf{2013}, \emph{110}, 046601.

\bibitem{WangJing_APE2014}
W.~Jing, L.~Zijing, and C.~Kwok~Sum, \emph{Appl. Phys. Express}  \textbf{2014},
  \emph{7}, 125102.

\bibitem{YuZhizhou_Carbon2016}
Z.~Yu, F.~Xu, and J.~Wang, \emph{Carbon}  \textbf{2016}, \emph{99}, 451.

\bibitem{ZhaiXC_PRB2016}
X.~Zhai, W.~Gao, X.~Cai, D.~Fan, Z.~Yang, and L.~Meng, \emph{Phys. Rev. B}
  \textbf{2016}, \emph{94}, 245405.

\bibitem{ZhaiXC_NJP2017}
X.~Zhai, S.~Wang, and Y.~Zhang, \emph{New J. Phys.}  \textbf{2017}, \emph{19},
  063007.

\bibitem{XuYong_CPB}
Y.~Xu, \emph{Chin. Phys. B}  \textbf{2016}, \emph{25}, 117309.

\bibitem{WangJiansheng_PRB2010}
E.~C. Cuansing and J.-S. Wang, \emph{Phys. Rev. B}  \textbf{2010}, \emph{81},
  052302.

\bibitem{Vincent2016}
V.~Michaud-Rioux, L.~Zhang, and H.~Guo, \emph{J. Comput. Phys.}  \textbf{2016},
  \emph{307}, 593.

\bibitem{Kajiwara_Nat2010}
Y.~Kajiwara, K.~Harii, S.~Takahashi, J.~Ohe, K.~Uchida, M.~Mizuguchi,
  H.~Umezawa, H.~Kawai, K.~Ando, K.~Takanashi, S.~Maekawa, and E.~Saitoh,
  \emph{Nature}  \textbf{2010}, \emph{464}, 262.

\end{thebibliography}

\newpage




\end{document}